\documentclass[twocolumn,superscriptaddress,aps,preprintnumbers,amsmath,amssymb,prl,nofootinbib]{revtex4-1}

\usepackage[titletoc,toc,title]{appendix}
\usepackage{titletoc}
\usepackage{titlesec}
\usepackage{physics}
\usepackage{textgreek}
\usepackage{amsmath}
\usepackage{amssymb}
\usepackage{amsfonts}
\usepackage{dsfont}
\usepackage{graphicx}
\usepackage{xcolor}
\usepackage{xfrac}
\usepackage{footmisc}
\usepackage{comment}
\usepackage{pifont}
\usepackage{times}
\usepackage[hidelinks]{hyperref}
\usepackage{,bm}
\usepackage{enumitem}
\usepackage{centernot}
\usepackage{cancel}
\usepackage{slashed}
\usepackage{relsize}
\usepackage{amsmath,amssymb,mathrsfs}
\usepackage{mdframed}
\usepackage{tcolorbox}
\usepackage{bm}
\usepackage{times}
\usepackage{epsfig}
\usepackage{verbatim}
\usepackage{bm}
\usepackage[utf8]{inputenc}
\usepackage{graphics}
\usepackage{graphicx,epsfig,amssymb,amsmath,color,cancel}
\usepackage{subfigure}
\usepackage[english]{babel}
\usepackage{adjustbox}
\usepackage{soul}
\usepackage[normalem]{ulem}

\definecolor{zima_blue}{HTML}{1393C1}
\hypersetup{setpagesize=false,bookmarksnumbered=true,bookmarksopen=true,%
colorlinks=true,linkcolor=zima_blue,urlcolor=zima_blue,citecolor=zima_blue,linktocpage=false}

\def\beq{\begin{equation}}
\def\eeq{\end{equation}}

\def\mc{\textcolor{blue}}

\newtcolorbox{mybox}[3][]{
  colframe = black,         %
  colback  = #2!10,
  coltitle = #2!20!black,
  title    = {#3},
  boxsep   = 2pt,           %
  top      = 1pt,           %
  bottom   = 1pt,           %
  #1,
}

\begin{document}

\title{Standard Model Baryon Number Violation at Zero Temperature from Higgs Bubble Collisions}

\author{Nabeen~Bhusal}
\email{nabeen.bhusal@desy.de}
\affiliation{Deutsches Elektronen-Synchrotron DESY, Notkestra{\ss}e 85, 22607 Hamburg, Germany}
\author{Simone~Blasi}
\email{simone.blasi@desy.de}
\affiliation{Deutsches Elektronen-Synchrotron DESY, Notkestra{\ss}e 85, 22607 Hamburg, Germany}
\author{Martina~Cataldi}
\email{martina.cataldi@desy.de}
\affiliation{Deutsches Elektronen-Synchrotron DESY, Notkestra{\ss}e 85, 22607 Hamburg, Germany}
\affiliation{II. Institute of Theoretical Physics, Universit\"{a}t Hamburg, 22761, Hamburg, Germany}
\author{Aleksandr~Chatrchyan}
\email{aleksandr.chatrchyan@su.se}
\affiliation{Nordita, KTH Royal Institute of Technology and Stockholm University, Hannes Alfv\'ens v\"ag 12, SE-106 91 Stockholm, Sweden}
\affiliation{The Oskar Klein Centre for Cosmoparticle Physics, Department of Physics, Stockholm University, AlbaNova, 10691 Stockholm, Sweden}
\author{Marco~Gorghetto}
\email{marco.gorghetto@desy.de}
\affiliation{Deutsches Elektronen-Synchrotron DESY, Notkestra{\ss}e 85, 22607 Hamburg, Germany}
\author{G\'{e}raldine~Servant}
\email{geraldine.servant@desy.de}
\affiliation{Deutsches Elektronen-Synchrotron DESY, Notkestra{\ss}e 85, 22607 Hamburg, Germany}
\affiliation{II. Institute of Theoretical Physics, Universit\"{a}t Hamburg, 22761, Hamburg, Germany}

\begin{abstract}
We compute for the first time baryon number violation at zero temperature  from Higgs bubble collisions and find that it can be of the same order as that from thermal sphalerons in the symmetric phase at electroweak temperatures. We study the dependence of the rate of Chern--Simons number transitions on the shape of the scalar potential and on the Lorentz factor of the bubble walls at collision via large-scale (3+1)D lattice simulations of the Higgs doublet and $SU(2)$ gauge fields. We estimate the resulting baryon asymmetry assuming some  CP-violating source activated by the Higgs-field variation during the phase transition.

\end{abstract}

\preprint{DESY-25-120}

\maketitle

{\bf Introduction--}
Baryon number ($B$) is a global symmetry of the Standard Model (SM) whose violation has never been observed experimentally. It is however %
efficiently violated by the chiral anomaly at high temperature $T$ through  $SU(2)$ vacuum transitions producing integer Chern--Simons (CS) number variations $\Delta N_{\rm CS}\propto \Delta B$, the %
sphaleron processes, which play a key role  in essentially all models of baryogenesis, whether at the electroweak scale \cite{Kuzmin:1985mm} or well beyond, as in leptogenesis \cite{Fukugita:1986hr,Davidson:2008bu}. 
These transitions can also occur at zero temperature when the Higgs field is driven far from equilibrium, e.g. if the SM electroweak phase transition (EWPT) is quenched, i.e. happens much faster than when thermally induced. This has been exploited in models of (local) {\it cold} baryogenesis where inflation takes place at the EW scale and the inflaton coupling to the Higgs quenches the Higgs mass parameter, leading to a tachyonic instability that triggers $\Delta N_{\rm CS}\neq 0$ at every point in space \cite{Garcia-Bellido:1999xos,Krauss:1999ng,Cornwall:2000eu,Copeland:2001qw,Garcia-Bellido:2003wva,
Enqvist:2010fd}. The rate of $B$ violation  from such a quenched EW crossover (i.e. not first-order EWPT) has been studied on the lattice in \cite{Smit:2002yg,Tranberg:2003gi,vanderMeulen:2005sp,Tranberg:2006ip,Tranberg:2006dg,Tranberg:2010af,Tranberg:2012jp,Mou:2015aia,Mou:2017atl,Mou:2017zwe,Mou:2017xbo,Mou:2018xto}.

In this letter, we compute for the first time the rate of $B$ violation at $T=0$ arising from Higgs bubble collisions. We show that this new source of $B$ violation can be of the same order as the one from thermal sphalerons. This opens up the possibility for a new mechanism of EW baryogenesis  in models where the reheat temperature associated with a first-order EWPT never exceeds the 130 GeV sphaleron freeze-out temperature~\cite{DOnofrio:2014rug}. 
 In this context, EW baryogenesis could be potentially realised even for very strongly-first order supercooled EWPT i.e. for bubble wall velocity exceeding the speed of sound, a regime where the standard non-local EW baryogenesis charge transport mechanism is highly suppressed \cite{Dorsch:2021ubz}. 
 Such frameworks are particularly motivated in Composite Higgs models where a light dilaton drives the EWPT \cite{Konstandin:2011dr,Konstandin:2011ds,Servant:2014bla,vonHarling:2016vhf,Bruggisser:2018mrt,Bruggisser:2022rdm,Bruggisser:2022ofg}. 
 See also Refs.\,\cite{Baldes:2021vyz,Azatov:2021irb,Dichtl:2023xqd,Cataldi:2024pgt} for baryogenesis with ultra-relativistic walls but additional sources of $B$ violation. More generally, this new source of $B$ violation may impact common models of EW baryogenesis as a new source of $B$ washout.
\begin{figure*}[t!]
{\includegraphics[width=5.35cm]{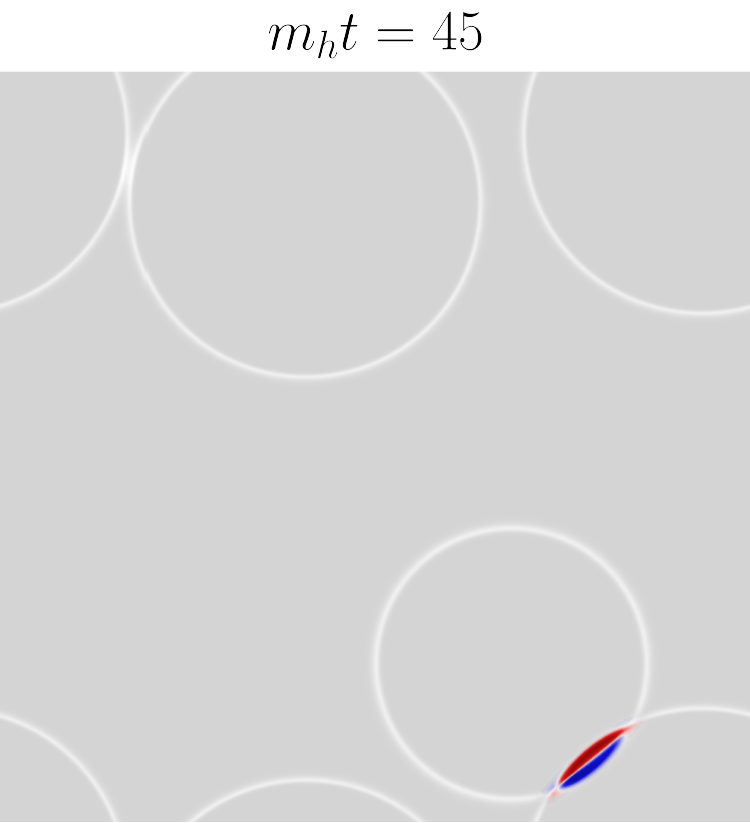}}
{\includegraphics[width=5.35cm]{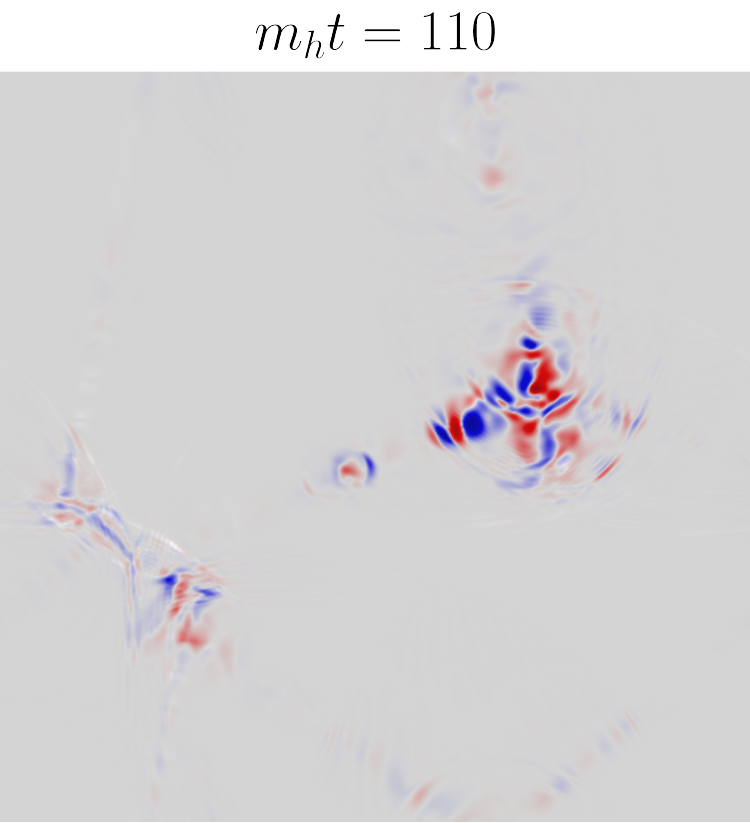}}
{\includegraphics[width=7cm]{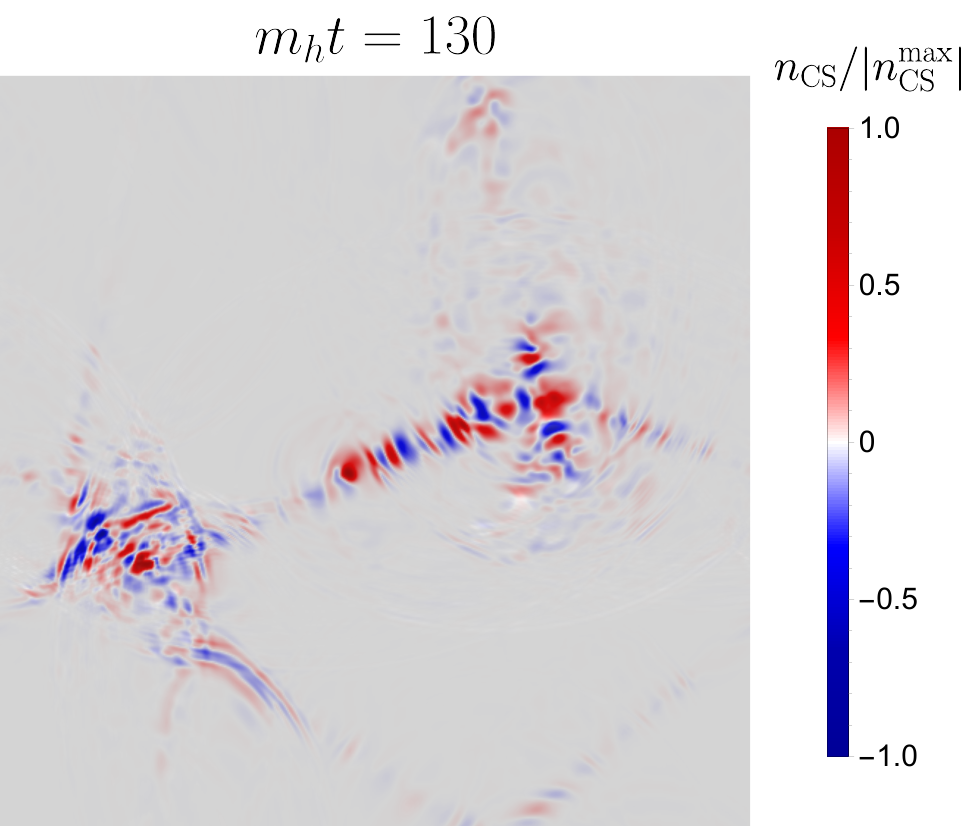}}
\caption{Snapshots of a two-dimensional slice of the Chern--Simons number density $n_{\rm CS}$ (red/blue) produced in our (3+1)D simulations from the collision of Higgs bubble walls (white), defined by $N_{\rm CS}(t)=\int d^3x \, n_{\rm CS}(t,x)$,  where $
n_{\rm CS}=({g^2}/{16\pi^2})\epsilon^{ijk}\,{\rm Tr}\left[W_i W_{jk}+\frac23i gW_{i}W_{j}W_{k} \right]$, for the potential degeneracy parameter $\epsilon=0.32$, normalized to its  maximum value over the slice $n^{\rm max}_{\rm CS}(t)$. See \href{https://www.youtube.com/watch?v=LhZFCxJ5-4g}{here} for an animation.}
\label{fig:bubble-snapshots}
\end{figure*}

The production of CS number in this work 
can be attributed
to the dynamics of the electroweak textures. 
These are configurations where the scalar field is in the vacuum manifold $M=SU(2)\times U(1)/U(1)$ everywhere in space, but defines a nontrivial map from $M$ onto the three-sphere $S^3= \mathbb{R}^3\cup\{\infty\}$, and are classified by third homotopy group $\pi_3(M)=\mathbb{Z}$. Thus, the energy of a texture is entirely due to the field gradient. 
In this paper, we consider $SU(2)$ as a simplified version of the SM gauge group, i.e. we set the hypercharge gauge coupling $g'=0$, as this captures the main physics of interest.
Textures can arise from the different $SU(2)$-phase orientations of the Higgs doublet $\phi$ across space after spontaneous symmetry breaking. %
A texture is labeled by the (integer) Higgs winding number 
$    N_{W}= \frac{1}{24 \pi^2} \int d^3 x \, \epsilon_{ijk} \text{Tr} (\partial_i \Phi) \Phi^{\dagger}  (\partial_j \Phi)\Phi^{\dagger} (\partial_k \Phi)\Phi^{\dagger}$, where $  \Phi= (i \sigma_2 \phi^*, \phi)/|\phi|^2 \in SU(2)
$, and the gauge field Chern--Simons number 
$N_{\rm CS}(t)-N_{\rm CS}(0) = \frac{g^2}{16 \pi^2} \int_0^t dt'\int d^3 x \, \text{Tr} [W^{\mu \nu} \tilde{W}_{\mu \nu}]$, where the dual of 
the $SU(2)$ field strength $W^{\mu \nu}$ is $\tilde{W}_{\mu \nu}= \frac{1}{2} \epsilon_{\mu \nu \rho \sigma} W^{\rho \sigma}$. A prototypical texture  with $N_W=1$  has $W_{\mu \nu}=0$ and  $\phi(x) = e^{- i \eta(r) x_i\sigma_i/r}(0,v)$, where $\eta(r)=\pi \tanh(r/r_t)$%
, $\sigma_i$ are Pauli matrices, $v=246\,$GeV %
is the Higgs expectation value and $r_t$ is the radius.
The minimal-energy (vacuum) configuration has $N_{\rm CS} = N_{W}$, while a texture has $N_{\rm CS}-N_W \neq 0 $. 
Textures with characteristic length scale  smaller than the inverse $W$ mass $m_W^{-1}$ decay 
by changing $N_W$, while those larger than $m_W^{-1}$ get `dressed' by the gauge fields and decay by altering %
$N_{\rm CS}$~\cite{Turok:1990zg}. 
Thus, crucially, the formation and subsequent decay of $SU(2)$-textures can  lead to a change in the CS number $\Delta N_{\rm CS}$, which eventually leads to baryon number production via the quantum anomaly,
$    \Delta B = N_F \Delta N_{\rm CS}$, where $N_F=3$ is the number of flavours.

Higgs bubble collisions generate Higgs inhomogeneities that produce $N_W$ and $N_{\rm CS}$, possibly directly or through formation and decay of textures. Our simulations indeed observe such %
production%
, see Fig.~\ref{fig:bubble-snapshots}. %
To quantify the change in the CS number, we define the CS number diffusion rate 
\begin{equation}
\label{eq:csrate_equ}
    \Gamma_{\rm CS} = \frac{1}{L^3} \frac{d \Delta N_{\rm CS}^2(t)}{dt} \equiv  \frac{d \Delta n_{\rm CS}^2(t)}{dt}\, ,
\end{equation}
with Chern-Simons variance
\begin{equation}
\label{eq:CSvariance}
    \Delta N_{\rm CS}^2(t) \equiv \, \expval{N^2_{\rm CS}(t)}-\expval{N_{\rm CS}(t)}^2 
\end{equation}
where $\langle . \rangle$ is the statistical average over realizations and $L^3$ is the volume. Notice that $\langle N_{\rm CS} \rangle =0$%
, as we do not include sources of CP violation.
It will be useful to compare Eq.\,(\ref{eq:csrate_equ}) with the SM thermal sphaleron rate in the symmetric phase~\cite{Moore:1998mh,Moore:2000ara,DOnofrio:2014rug},
$\Gamma_{\rm th}  \sim 25 \alpha_w^5 T^4$,
and with the rate from a quenched EW crossover~\cite{Garcia-Bellido:1999xos,Tranberg:2003gi,vanderMeulen:2005sp},
$\Gamma_{\rm tach} \equiv \alpha_w^4 T_{\rm eff}^4$,  where $T_{\rm eff}$  parametrizes the efficiency of baryon number violation induced by the tachyonic instability and depends on the amount of quenching of the EW crossover.
\begin{figure*}[t!]
\includegraphics[width=0.54\linewidth]{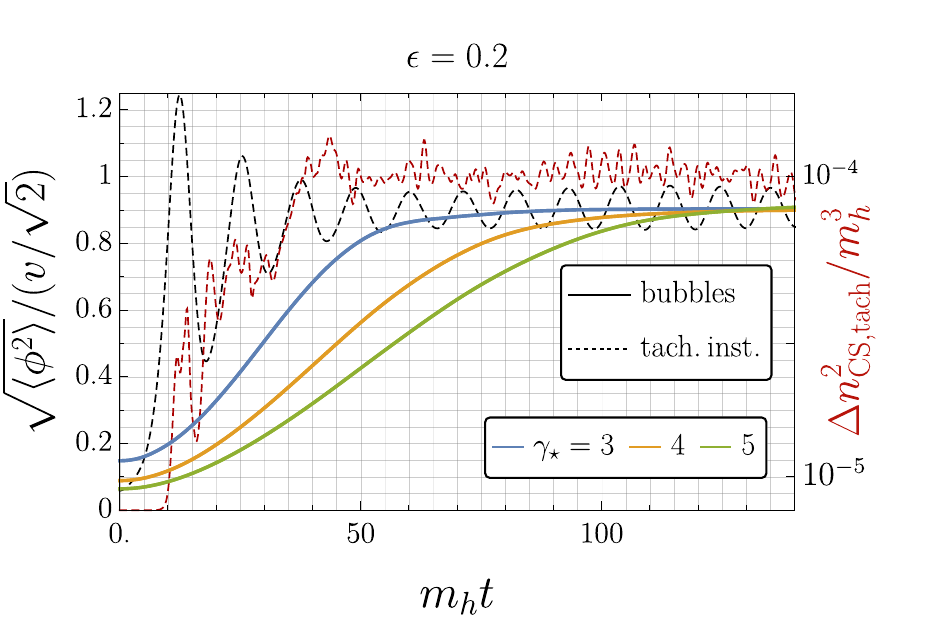}\hfill
\includegraphics[width=0.46\linewidth]{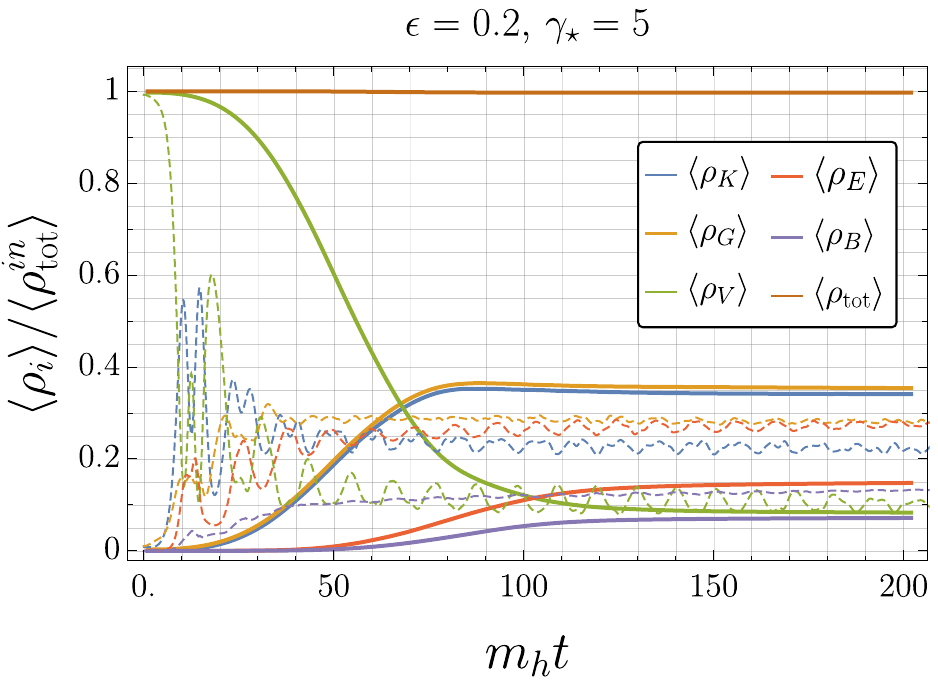}\\[-1em]
\caption{
Solid lines indicate bubble collisions, whereas dashed lines represent a typical tachyonic transition with an instantaneous quench (Higgs field rolling down its $T=0$ potential from the origin). 
Left: Time evolution of volume-averaged Higgs norm during the phase transition in units of its value today $v=246$ GeV, for different values of Lorentz-factor $\gamma_\star$ of the walls at collision. The dashed red line is the CS variance for the tachyonic transition. Right: Evolution of the energy density components normalized by the total initial energy density $\rho_{\rm tot}^{in}$.
The scalar kinetic and gradient, and $SU(2)$ electric and magnetic energies are: $\rho_K  =  |\dot{\phi}|^2 \, , \hspace{1mm} \rho_G = |D_i \phi|^2\,, \hspace{1mm}   \rho_E  = \frac12W_{0i}W_{0i}\,, \hspace{1mm}  \rho_B  = %
W_{jk}^a W_{jk}^a\,$; $\rho_V$ is the potential. We 
use a degeneracy parameter $\epsilon=0.2$, and average over 30 simulations in both panels.
}
\label{fig:bubblecoll_higgs_norm_1}
\end{figure*}

{\bf Real-time simulations--} We rely on SM-physics only, except that we assume a first-order EWPT triggered by a new light scalar, with the Higgs field tracking its bubble profile. For simplicity, however, we model the transition using a modified single-field Higgs potential in which $\phi = 0$ is a local minimum, 
see Suppl. Mat.\,\ref{sec:Appendix-Lattice}. We solve the Higgs and gauge boson equations of motion numerically on a periodic cubic grid with $N_x^3 \in [500^3 - 1500^3]$ points, setting $m^2_W/m^2_h=0.4$ as in the SM (neglecting hypercharge). We expect our main results on the CS rate to be barely affected when the full $SU(2) \times U(1)$ 
is considered. The main model-dependent parameter is the degeneracy parameter $\epsilon$, which characterizes the scalar potential shape, and plays an important role in the field dynamics during and after bubble collisions~\cite{Konstandin:2011ds,Jinno:2019bxw,Gould:2021dpm}. It is defined by 
  $  \epsilon %
  = \frac{V_{\rm barr}}{V_{\rm barr} +|\Delta V|}\in (0,1)$,
 where $V_{\rm barr}$ is the barrier height and %
 $\Delta V$ is the energy difference between the $\phi=0$ and $\phi=v/\sqrt{2}$ vacua~\cite{Jinno:2019bxw}%
 . We have
$\epsilon \to 1$ for  degenerate vacua, $|\Delta V| \ll V_{\rm barr}$,
and $\epsilon \to 0$ for non-degenerate vacua, $|\Delta V| \gg V_{\rm barr}$. For large $\epsilon$, the walls tend to reflect upon collision, temporarily recreating a region of false vacuum that eventually shrinks after a several iterations.
 Instead, for  $\epsilon\ll1$ the barrier is not effective in trapping the field in the false vacuum, and the true minimum is eventually reached after just a few oscillations. 

We nucleate $N_b = 10$ bubbles simultaneously, but verify that our results are independent of $N_b$, see Suppl. Mat.\,\ref{sec:Appendix-Lattice}. %
Their profile $h_c(r)$ is the $O(4)$ solution of the equations of motion characterized by a critical radius $R_c$ and wall width $l_0$, with  $R_c>l_0  \sim m_h^{-1}$. 
Contrary to tachyonic instability studies, no (quantum) fluctuations are introduced. %
Bubbles are nucleated with a random position and orientation $\theta_i$ in the $SU(2)$ space: %
$\phi = \frac{1}{\sqrt{2}} e^{- i \theta_i \sigma_i}(0,h_c(r))\, .$
The different orientations allow the scalar field to acquire a nonzero winding number, generating a current that ultimately sources the gauge fields (initially set to zero): 
$D_\mu W^{\mu \nu} = 2 g \,\text{Im}\,(\phi^\dagger D^\mu \phi)$.
After collision, bubbles fill the whole volume with the true vacuum%
. Defining the bubble radius at collision $R_\star = \gamma_\star R_c$, we estimate the Lorentz factor of the wall at collision, $\gamma_\star$, via $L^3 \sim N_b \left(2/{\sqrt{3}}\right)^3 (4\pi/3) \left(R_c \gamma_\star \right)^3$. Fig.~\ref{fig:bubblecoll_higgs_norm_1}-left shows the time evolution of the volume-averaged Higgs norm during the transition, for varying $\gamma_\star$. 
Our analysis applies to EW bubbles that collide in the runaway regime, see  Sec.\,\ref{sec:runaway} of the End Matter. %

The evolution of the volume-averaged energy density components normalized by the initial total energy $\rho_{\rm tot}^{in}$ is shown in Fig.\,\ref{fig:bubblecoll_higgs_norm_1}-right ($\rho_{\rm tot}$ is conserved). 
Most of the potential energy is converted into Higgs kinetic and gradient energy, with a non-negligible fraction transferred to electric and magnetic fields. This picture does not change notably for different values of $\epsilon$ from what shown in Fig.\,\ref{fig:bubblecoll_higgs_norm_1}.%

Non-zero $N_W$ and $N_{\rm CS}$ are produced as  the bubbles collide (see Fig.~\ref{fig:bubble-snapshots}). %
Once the phase transition completes, $N_{\rm CS}$ relaxes toward $N_W$ in the vacuum configuration; see Fig.\,\ref{fig:totwn_csn_1} in the End Matter for an example.
Note that the collision of only two bubbles does not produce $N_{W}$ due to the cylindrical symmetry. %
The CS variance $\Delta n^2_{\rm CS,tach}$ in the original cold baryogenesis from a tachyonic instability with instantaneous quench  is shown for comparison in Fig.\,\ref{fig:bubblecoll_higgs_norm_1} left.

{\bf Dependence of Chern-Simons rate on $\gamma_\star$ and $\epsilon$ -- }
We study the dependence of the CS rate, $\Gamma_{\rm CS}$, on the EWPT parameters: i) $\gamma_\star$, the average Lorentz-boost factor of the bubble walls at collision, which in a cosmological context is proportional to $\beta/ H $, where $\beta^{-1}$ is the phase transition  duration and $H$ is the Hubble parameter; ii) the degeneracy parameter $\epsilon$, encoding the potential shape and the dynamics of bubble wall collisions.
We perform simulations varying %
$\gamma_\star$ (via changing $L$ and thus $R_\star$), and for several values of $\epsilon$, with %
$N_b=10$ and %
$l_\star/dx \gtrsim 2$ points per bubble wall  (see Suppl. Mat.~\ref{sec:Appendix-Lattice} for details). The statistical average in Eq.\,\eqref{eq:CSvariance} is done  over 30 independent simulations. Our results for $\Gamma_{\rm CS}$ 
have to be extrapolated to the physical limit of large $\gamma_\star$, or small $\beta/H$,
for which there is a large hierarchy between the bubble size and the microscopic scale set by the Higgs mass.
We are able to test values of $\gamma_\star\lesssim8$, and since  $ H R_\star= \gamma_\star H R_c = (8 \pi)^{1/3} v_w (\beta/H)^{-1}$
with $H\sim v^2/M_{\rm Pl}$ and $v_w \sim 1$ the wall velocity, %
this corresponds to
$\beta / H \gtrsim 10^{15}$, which are much larger than the typical ones for a strong first-order PT, namely $\beta / H \sim 10^1-10^{3}$.

\begin{figure*}[t!]
\includegraphics[width=8.4cm]{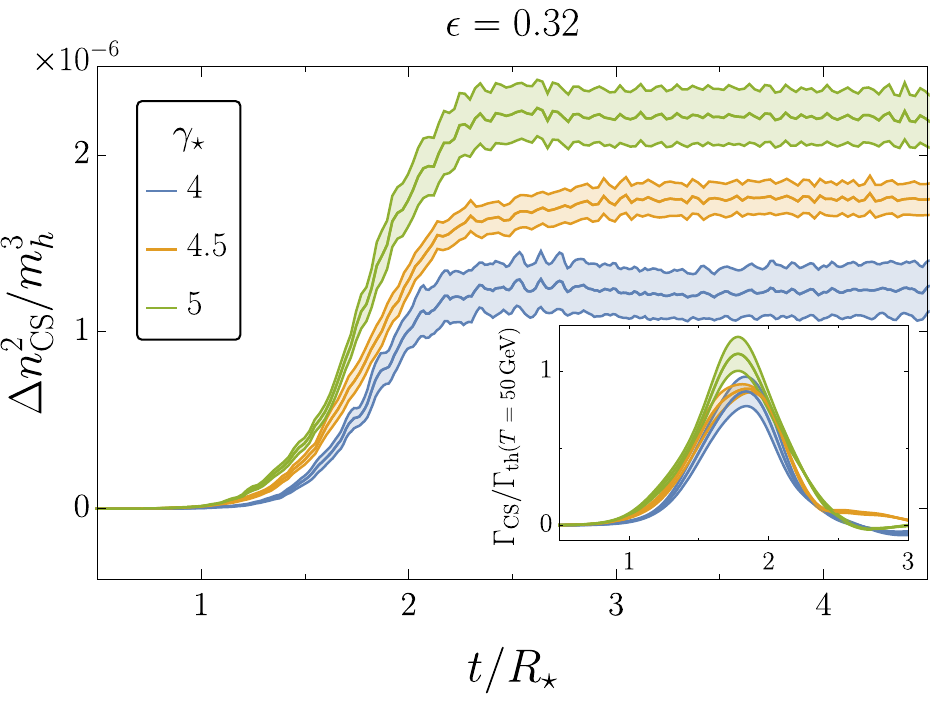}
\hspace{2mm}
\includegraphics[width=9.1cm]{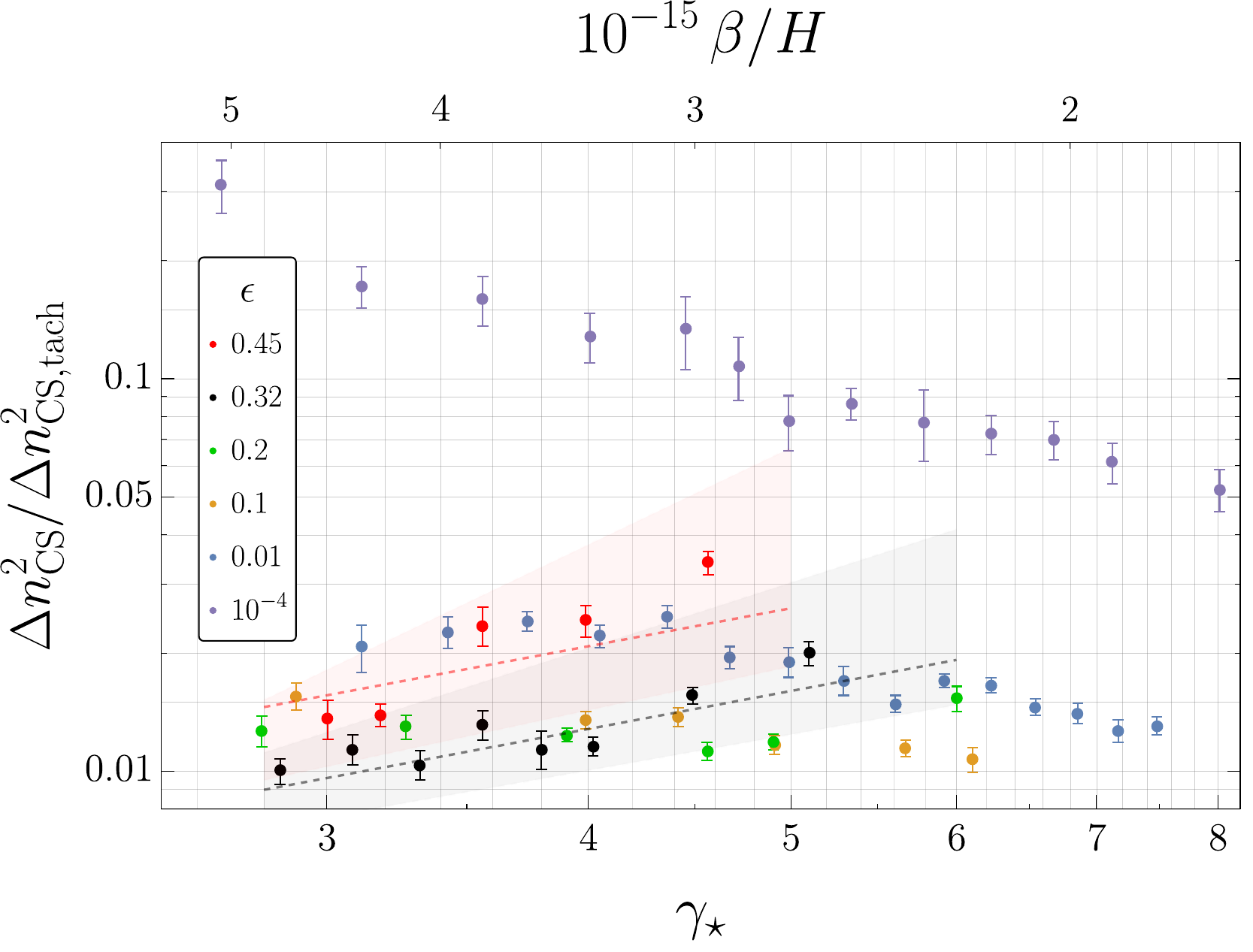}
\caption{Left: Evolution of Chern--Simons variance $\Delta n_{\rm CS}^2 $ and rate  $\Gamma_{\rm CS} $ as a function of  $\gamma_\star$, for $\epsilon =  0.32$. %
For each $\gamma_\star =\{4,4.5,5\}$, we have $m_h  R_\star = \{ 80, 90, 100\}$. 
Right: {Asymptotic value of CS variance $\Delta n_{\rm CS}^2$ as a function of the Lorentz-boost factor $\gamma_\star$ at collision, for different values of $\epsilon$, normalized by the CS variance from a tachyonic instability,  
$\Delta n_{\rm CS,tach}^2=10^{-4} m_h^{3}$\,\cite{Tranberg:2006dg,Mou:2017atl}. While for $\epsilon\ll1$ the CS variance appears to decrease with $\gamma_\star$, for $\epsilon \gtrsim 0.3$ there is some evidence that it increases $\propto \gamma_\star^a$ with $a\gtrsim 1$ (e.g. black and red). %
This corresponds to a non-decreasing $\Gamma_{\rm CS}$ with respect to $\gamma_\star$ (inset in the left panel) and hints at a large baryon number violation at the physical point $\gamma_\star \gg 1$. %
The bands denote the 2$\sigma$ error on the fitted exponent $a$; for comparison, the dashed line corresponds to $a = 1$, still consistent with the data. 
}
}
\label{fig:var_eps02}
\end{figure*}

In Fig.~\ref{fig:var_eps02}-left we show the time evolution of the CS variance, $\Delta n_{\rm CS}^2$, and the CS rate $\Gamma_{\rm CS}^{\rm}$, defined by Eqs.~\eqref{eq:csrate_equ},~\eqref{eq:CSvariance}, for $\epsilon= 0.32$. At a fixed $\epsilon$, the CS variance may be parametrized by 
\begin{equation}
\label{eq:estimateCS}
\Delta n_{\rm CS}^2 = c \cdot \gamma_\star^a f(t/R_\star), 
\end{equation}
where $c$ is a constant, the exponent $a$ encodes the possibility of a dependence on $\gamma_\star$, either increasing or decreasing%
, $f$ is the profile observed in Fig.\,\ref{fig:var_eps02}. %
In a timescale set by the completion of the EWPT,  ${\cal O}(R_\star$), the CS variance has grown to a sizable value $\Delta n^2_{\rm CS}/m_h^3 \sim 10^{-6}$. %
This value corresponds to $\Gamma_{\rm CS}$ being roughly of the same order as the thermal sphaleron rate in the symmetric phase, $\Gamma_{\rm th}$, at $T \sim 50\,\text{GeV}$, active for the whole duration of the transition, see inset of Fig.~\ref{fig:var_eps02}-left. Despite the small range of $\gamma_\star$ considered for this value of $\epsilon$, as we discuss next, we see some indication for a dependence compatible with $a=1$ in Eq.\,\eqref{eq:estimateCS}. Since $\Gamma_{\rm CS} \propto \gamma_\star^{a-1}$, this would suggest that the (peak of the) CS rate approaches a constant at $\gamma_\star\gg1$ for this large  $\epsilon$, see the inset. 
The characteristic timescale associated with $B$ violation, when $\Gamma_{\rm CS} \neq 0$ and reaches macroscopic values, is $\sim R_\star$  (i.e. fractions of Hubble time), in contrast with the tachyonic instability in original cold baryogenesis %
where the relevant timescale is the time of equilibration of the Higgs modes, which is microscopic.

In Fig.\,\ref{fig:var_eps02}-right we show the asymptotic value of $\Delta n_{\rm CS}^2$ after the EWPT completes as a function of $\gamma_\star$ for different  $\epsilon$, normalized to the variance from a tachyonic phase transition, $\Delta n^2_{\rm CS,tach}$, shown in Fig.\,\ref{fig:bubblecoll_higgs_norm_1} for comparison. %
For small $\epsilon \lesssim 0.1$, $\Delta n_{\rm CS}^2$ decreases with $\gamma_\star$ (or equivalently $R_\star$), namely $a <0$ in Eq.\,\eqref{eq:estimateCS}. For $\epsilon = 10^{-4}$ the numerical results are compatible with $a=-1$. This dependence is consistent with CS production being purely a surface effect, which vanishes as $R_\star^{-1}$ in the physical limit of macroscopic bubbles (with $N_{\rm CS}$ produced only around the collision points). On the other hand, larger values of $\epsilon \gtrsim 0.2$ suggest a CS variance that becomes constant, $a=0$, or grows, $a>0$, in the physical limit of $m_h R_\star\to \infty$ and $\gamma_\star\rightarrow \infty$, as noticed before for $\epsilon = 0.32$. 
In the tested dynamical range, the CS variance from bubble collisions is smaller than the one from tachyonic instability with instantaneous quench by one or two orders of magnitude. %
However, we stress again that these results need to be extrapolated to the physical point with $\gamma_\star \gg 1$.

The regime $\epsilon\gtrsim0.2$, where interestingly $N_{\rm CS}$ does not appear to decrease with $\gamma_\star$, is challenging to study numerically at $\gamma_\star \gg 1$ because $R_c$ is parametrically larger than the wall width. 
Since we only test a relatively small range of $\gamma_\star$ and results have still quite large statistical uncertainties, the data does not allow us to reconstruct $\Delta n_{\rm CS}^2$ reliably at macroscopic $\gamma_\star$. 
In any case, for $\epsilon \gtrsim 0.3$ the data favors a growing variance $\Delta n_{\rm CS}^2 \propto \gamma_\star^{a}$, with  $a \gtrsim 1$. The best-fit values are $a \simeq 1.3(2)$ and $a \simeq 1.8(4)$ for $\epsilon = \{0.32,0.45\}$ respectively.  Taken together, these results suggest that $\Gamma_{\rm CS}\propto\gamma_\star^{a-1}$ is likely not volume-suppressed at large $\gamma_\star$.
Stronger evidence for this trend, possibly suggesting the value $a\simeq1$, appears in the (1+1)D $U(1)$ gauge-field case (see Fig.\,\ref{fig:1D_spectrum} and the discussion in Sec.\,\ref{app:results_1D} of the End Matter), 
which probes larger $\epsilon$ and $\gamma_\star$. The numerical treatment of such (1+1)D system is discussed in more detail in Suppl. Mat.~\ref{app:U(1)}. %
Despite these limitations, to the best of our knowledge, this work represents the current state of the art in real-time (3+1)D simulations of first-order PT in $SU(2)$ gauge theory (simulations of 2-bubble collisions with(out) $U(1)$-gauge field have been carried out up to $\gamma_\star \sim 5$ in Ref.\,\cite{Lewicki:2020azd} ($\gamma_\star = 20$ in Ref.\,\cite{Gould:2021dpm}), while multiple bubble collisions without gauge fields were simulated in (3+1)D with $\gamma_\star$ as large as $\gamma_\star \simeq 6$ in \cite{Cutting:2018tjt} and $\gamma_\star \simeq 50$ in (1+1)D in \cite{Jinno:2019bxw}).%

Our main conclusion is that $\epsilon$ is a crucial parameter determining whether $B$ can be produced over a large volume. 
If the vacua are highly non-degenerate (small $\epsilon$), bubble collision is limited to the surfaces. 
Instead, for nearly degenerate vacua (large $\epsilon$), walls pass through each other, extending the collision region and the production of $N_{\rm CS}$ far out from the initial point. 
CS transitions then
extend to a sizable fraction of the whole volume due to the subsequent transfer of energy from the wall kinetic energy to other gauge and Higgs field configurations  (see third panel of Fig \ref{fig:bubble-snapshots}).

{\bf Chern--Simons variance from the spectrum--} The CS number density spectrum $P_{\rm CS}(k)$ provides a more detailed information on the length scales at which the CS number is produced. This is defined by $\langle n_{\rm CS}(k)n_{\rm CS}^*(k')\rangle= (2\pi)^3 \delta^3(k-k')\frac{2\pi^2}{k^3}P_{\rm CS}(k)$, where $n_{\rm CS}(k)=\int d^3x e^{-ikx}n_{\rm CS}(x)$ and $N_{\rm CS}=\int d^3x\,n_{\rm CS}$.  Since the field fluctuations %
are uncorrelated beyond the typical bubble separation, we expect $P_{\rm CS}(k)$ to have the  white-noise behavior  $P_{\rm CS}(k) = C k^3/(2\pi^2)$ for $k \ll 2\pi/R_\star$%
. From Eqs.\,\eqref{eq:csrate_equ} and \eqref{eq:CSvariance} one readily verifies that the coefficient $C$ is in fact the CS variance, $C = \Delta n_{\rm CS}^2$, see Suppl. Mat.~\ref{sec:Appendix-Lattice}. %

\begin{figure}[b!]
\centering
\includegraphics[width=9cm]{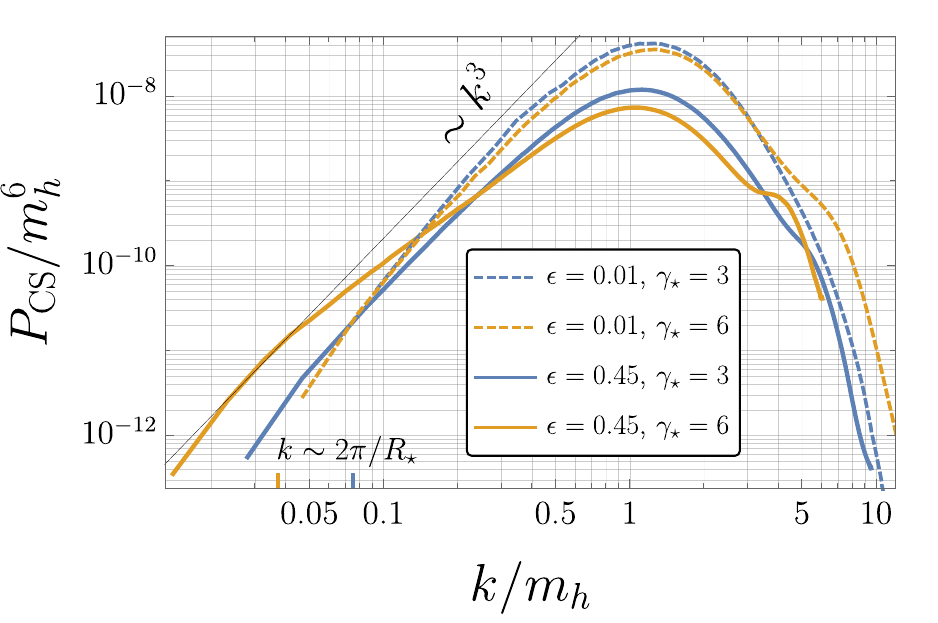}
\caption{Averaged CS number spectra for small  and large  $\epsilon$, and for $\gamma_\star=\{3,6 \}$. For $\epsilon=0.45$, the CS spectrum develops an IR shoulder at $k \sim 2\pi/R_\star$ (blue and orange ticks on the $x$-axis), with a $k^3$ white-noise behavior at smaller $k$.}
\label{fig:3D_spectra_big_eps_045}
\end{figure}

Fig.~\ref{fig:3D_spectra_big_eps_045} shows power spectra for $\epsilon = \{10^{-2},0.45\}$ and $\gamma_\star=\{3,6\}$. Most of the CS number is generated at scales of order $m_h^{-1}$, since both spectra peak at $k \simeq m_h$. 
However, for $\epsilon = 0.45$, the spectra also develop a small IR shoulder at $k \simeq 2\pi/R_\star$. This becomes increasingly pronounced as $\gamma_\star$ grows, at the expense of the $m_h$ peak, which is progressively reduced at larger $\gamma_\star$. At lower momenta, the spectrum roughly follows the expected white-noise behavior. These properties of the CS spectrum are much clearer in our (1+1)D simulations, as shown in Fig.\,\ref{fig:1D_spectrum} and further detailed in Suppl.Mat.\,\ref{app:U(1)}. Taken together, these results suggest that in the $\gamma_\star \gg 1$ limit, the CS spectrum peaks at $k \simeq 2\pi / R_\star$ for large $\epsilon$; consequently, the coefficient $C = \Delta n_{\rm CS}^2$ grows with $\gamma_\star$, and the production occurs over macroscopic scales $\sim R_\star$. The IR shoulder is instead absent for $\epsilon = 10^{-2}$, explaining the different behavior for the CS variance with $\gamma_\star$ in Fig.\,\ref{fig:var_eps02}-right.

{\bf Estimate for the baryon asymmetry --  }
This letter computes a universal effect of the EW theory, which can  have implications for a large range of Standard-Model extensions.
In contrast,  a baryon asymmetry calculation requires the introduction of CP-violating interactions and a model-dependent analysis. We can nevertheless argue how this new effect could lead to the correct baryon asymmetry.
 In a thermal environment, the baryon number density $n_b$ obeys a Boltzmann equation that combines the rate of $B$ violation $\Gamma$,  the effect of CP violation in producing a net asymmetry $\mu$, and washout terms $\mathcal{A}$:
$
    \frac{\partial}{\partial t} n_b(x,t) = \Gamma \left( \xi - \mathcal{A} \, n_b \right),
$
where $\xi \equiv \mu/T$ is the effective chemical potential for the baryons. 
$\Gamma$ and $\mu$ are in general space-time dependent, while we will neglect $\mathcal{A}$ for simplicity. In standard electroweak baryogenesis (EWBG) for instance, the chemical potential is non-zero only in the vicinity of the expanding bubble, while sphalerons are active only in the symmetric phase ahead of the wall, $\Gamma = \Gamma_{\rm th}$. 
On the other hand, the scenario of cold baryogenesis introduced in Ref.\,\cite{Garcia-Bellido:1999xos} does not rely on a first-order PT for driving the system out of equilibrium, but rather on the tachyonic instability at the top of the SM Higgs potential. This dynamics can be considered homogeneous in space, at least over scales that are much larger than the typical size of the SM textures\,\cite{Garcia-Bellido:2003wva}. 
The baryon asymmetry 
can be estimated by referring to a volume-averaged sphaleron rate and chemical potential. The former is
extracted from numerical simulations via  the diffusion of the CS number, Eq.\,\eqref{eq:csrate_equ}.
CP violation is usually introduced via the following dimension-six operator\,\cite{Garcia-Bellido:1999xos,Garcia-Bellido:2003wva,Tranberg:2006ip} (see however \,\cite{Tranberg:2009de} and \,\cite{Servant:2014bla} for other choices),
$\mathcal{O} = \frac{g^2}{32\pi^2\Lambda^2}\phi^\dagger \phi \, W_{\mu\nu}^a\tilde{W}^{\mu\nu \,a}$,
where $\Lambda$ corresponds to the new physics scale at which  $\mathcal{O}$ is generated. Current bounds on $\Lambda$ are dominated by those from the electron electric dipole moment, see e.g.~\cite{Cirigliano:2019vfc}, and lead to $\Lambda \gtrsim6$\,TeV.  
$\mathcal{O}$ generates an effective chemical potential $\mu = N_F^{-1} \frac{1}{\Lambda^2} \frac{d}{dt} \langle \phi^\dagger \phi \rangle$,
as can be seen by integrating by parts $\int d^4 x \,\mathcal{O}$ and taking volume averages, $\langle \dots \rangle$, see e.g.\,\cite{Dine:1991ck,Garcia-Bellido:1999xos}. 

The scenario considered in this paper is somewhat in between EWBG and the original cold baryogenesis setup, as it relies on a first-order EWPT but also on the subsequent dynamics leading to a re--equilibration of the Higgs and gauge fields, as advocated in \cite{Konstandin:2011ds,Servant:2014bla}.
We may estimate the baryon asymmetry following the approach adopted in cold baryogenesis. 
We stress however that this estimate is based on volume-average quantities, and it may differ significantly in case the CS transitions are localized around the (collided) walls. By considering the same source of CP violation from the operator $\mathcal{O}$, we can derive the effective chemical potential by referring to Fig.\,\ref{fig:bubblecoll_higgs_norm_1}--left and introducing the following parametrization for the Higgs average variance:
$\langle |\phi|^2 \rangle = \frac{v^2}{2} f(b \, t/R_\star)$,
where $b \sim \mathcal{O}(1)$ accounts for a possibly different time scale compared to the CS variance in Eq.\,\eqref{eq:estimateCS} (in fact, the Higgs norm converges faster than the CS variance).
The total baryon asymmetry is then given by
$n_b = \int dt \, \Gamma_{\rm CS}(t) {\mu(t)}/{T_{\rm eff}(t)}$, 
where $T_{\rm eff} = \Gamma_{\rm CS}^{1/4}/\alpha_w$. We then obtain for the baryon yield to entropy density ratio ${n_b}/{s} \simeq 10^{-10} \gamma_\star^{3(a-1)/4}\left(\frac{30\,{\rm GeV}}{T_{\rm rh}}\right)^3\left(\frac{10\,{\rm TeV}}{\Lambda}\right)^2 \,$, where $T_{\rm rh}$ is the reheat temperature after the EWPT. In deriving this expression we have taken $R_c = 10\, m_h^{-1}$ as a typical value, and used that $\frac{b}{2} \int_0^\infty d x f^\prime(x) f^\prime(b x) \approx 0.1$. %
This estimate implies that $a \geq 1$ is required for successful baryogenesis (i.e. $n_b/s\gtrsim 10^{-10}$); otherwise, $n_b/s$ is suppressed at large $\gamma_\star$. %
However, $a \geq 0$ may suffice if the CS transitions occur locally, following the motion of the walls after collision, as suggested by the (1+1)D case in Fig.\,\ref{fig:1D_spectrum} (bottom panel). 
In this case the previous estimate based on volume--average quantities would not apply, and the convolution of a local CS rate with the regions where CP violation occurs may still lead to unsuppressed baryon asymmetry, similarly to %
EWBG. %
The analysis of the (3+1)D system including a dynamical CP-violating operator (e.g., $\mathcal{O}$) is however needed to ultimately establish the viable parameter space (e.g., $\epsilon$ and $\Lambda$) for successful cold baryogenesis. For such analysis, the key quantity will be the (non--zero) average $\langle N_{\rm CS}\rangle$, which is directly related to baryon number, rather than the variance $\langle N_{\rm CS}^2\rangle$ as considered here.

{\bf Conclusion--}
SM Higgs bubble collisions provide a new sizeable source of Chern--Simons number. We calculated from first-principles (3+1)D real-time
simulations the corresponding CS rate in the runaway regime and showed its sensitivity on the shape of the scalar potential driving the first-order EWPT. For $\epsilon\gtrsim 0.3$, there is no volume suppression. The next step will be to implement CP violation on the lattice to evaluate  precisely the net baryon number. 
 This may provide an alternative realization of electroweak baryogenesis in the large bubble wall velocity limit, which does not rely
on the charge transport mechanism. More generally, it will be important to investigate  the impact of this potentially new washout effect on the standard EW baryogenesis mechanism. An interesting extension is to include the $U(1)$ gauge field and calculate magnetic-field production in the spirit of e.g.~\cite{Vachaspati:2024vbw,Di:2020kbw,Di:2024gsl,Di:2025ncl}. These simulations can also predict the resulting gravitational-wave background. %

{\bf Acknowledgements--} %
We thank Anders Tranberg and Hyungjin Kim for discussions. This research was supported in part through the Maxwell computational resources operated
at Deutsches Elektronen-Synchrotron DESY, Hamburg, Germany.
This work, and MC in particular, is supported by the Deutsche Forschungsgemeinschaft under Germany’s Excellence Strategy---EXC 2121 ``Quantum Universe"---390833306.
 NB acknowledges support by the Deutsche Forschungsgemeinschaft (DFG, German Research Foundation) under the DFG Emmy Noether Grant No. PI 1933/1-1. The work of MG is supported by the Alexander von Humboldt foundation. GS thanks the theory group at Berkeley for hospitality in the very last stages of completion of this work as well as the European Union’s Horizon Europe programme under Marie
Skłodowska-Curie Actions – Staff Exchanges (SE) grant agreement No 101086085 -ASYMMETRY.

\appendix

\begin{widetext}

\begin{center}
\textbf{\it\Large End Matter
}
\end{center}

\begin{figure}[h!]
{\includegraphics[width=7.4cm]{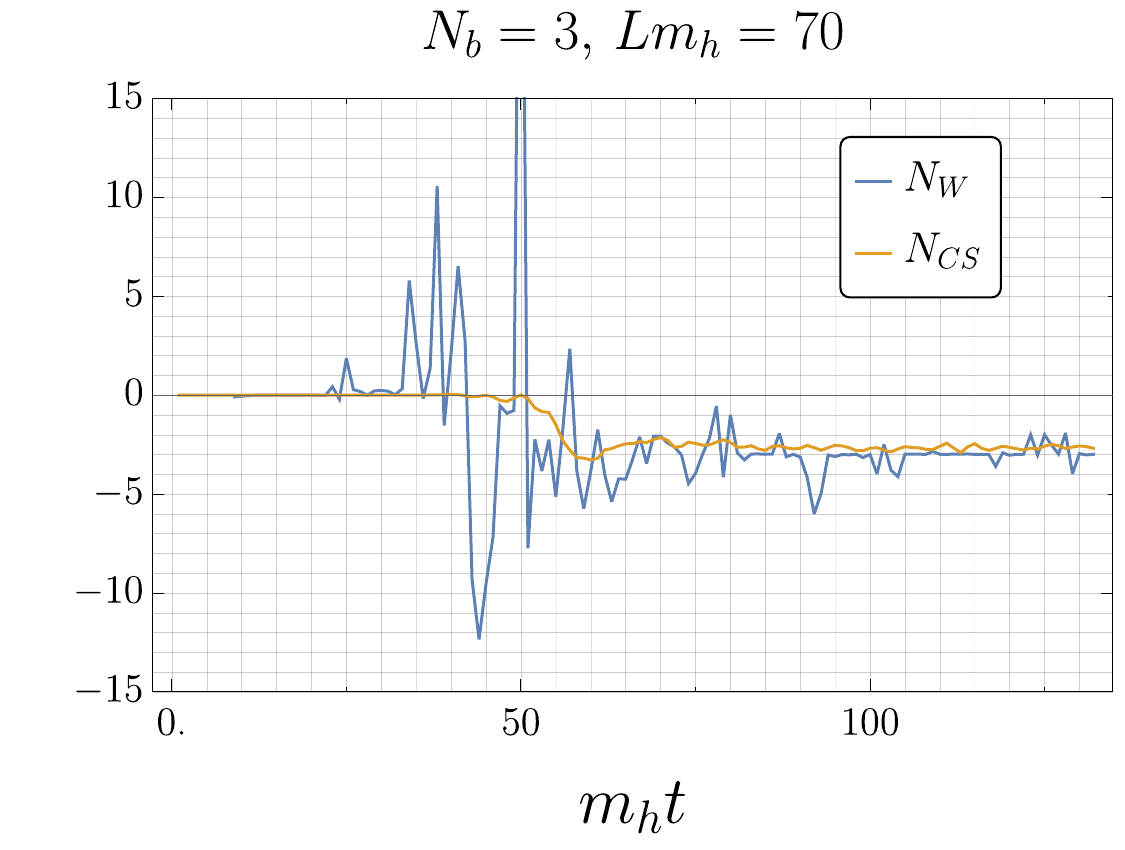}}
\hspace{2mm}
{\includegraphics[width=7.6cm]{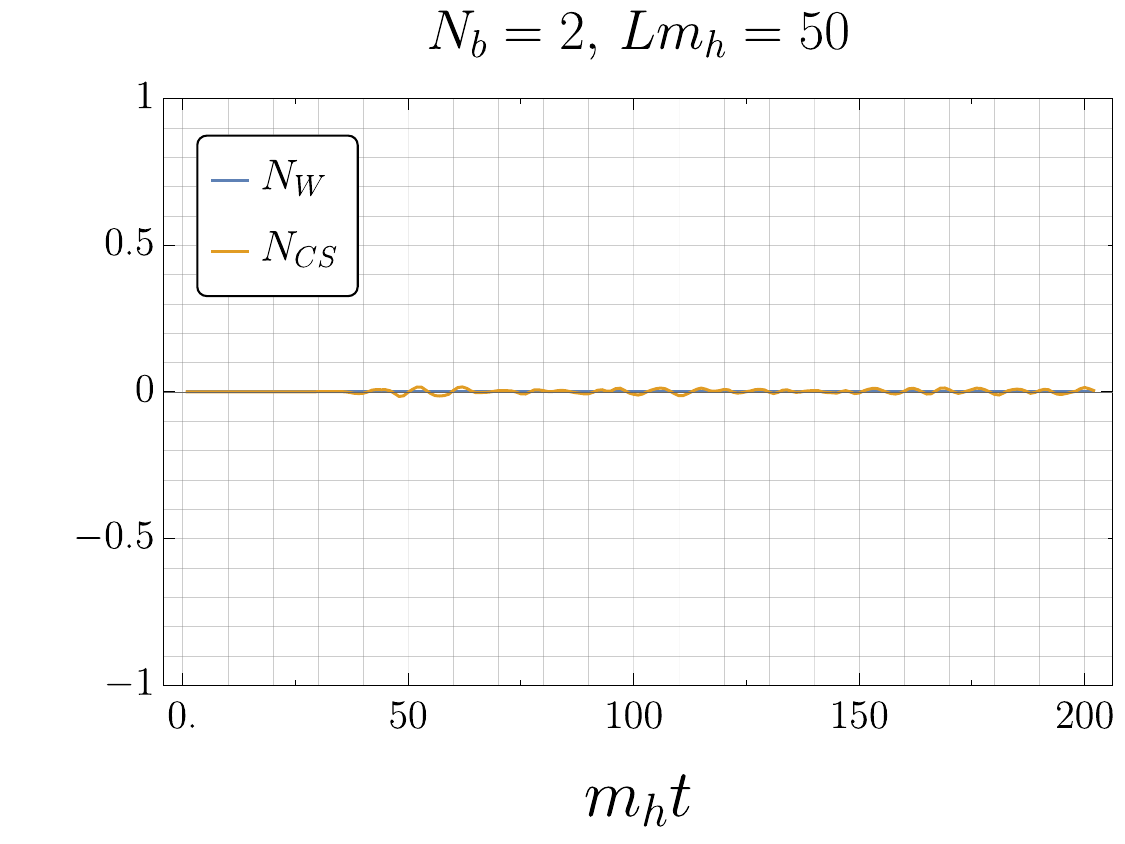}}
\caption{The evolution of the winding and Chern--Simons numbers $N_W$ and $N_{\rm CS}$ from a 3-bubble collision (left panel), and a 2-bubble collision (right panel) for $\epsilon=0.2$ in (3+1)D. or $N_b=3$, $N_W$ is produced as bubbles collide, while for $N_b=2$, $N_W \equiv 0$ because of the symmetry of the system. While locally $n_{\rm CS}$ is large as shown in the symmetric blue and red arches in the bottom-right two-bubble collision of the first snapshot of Fig.~\ref{fig:bubble-snapshots}, once integrated over the volume, $N_{\rm CS}$ is suppressed for $N_b=2$. In both cases, $N_{\rm CS}$ asymptotically relaxes to $N_W$ as expected in a vacuum configuration.} 
\label{fig:totwn_csn_1}
\end{figure}

\section{Winding and CS number}
\label{sec:checks}

In Fig.\,\ref{fig:totwn_csn_1} we show the evolution of the winding number and the CS number during the collision of three (left panel) and two (right panel) runaway bubbles. The two quantities converge at the end of the EWPT, as expected for vacuum configurations where indeed $N_{\rm CS} = N_W$. This represents a non-trivial cross check of our numerical simulations, and shows how winding configurations play a crucial role in triggering CS transitions. Bubble collisions initially generate a non-vanishing $N_W$, after which $N_{\rm CS}$ asymptotes toward $N_W$. %
However, $N_{\rm CS}$ does not coincide exactly with $N_W$, since part of the Chern--Simons number can be carried away by propagating $W$-boson waves. Conversely, we have checked that the collision of two bubbles leads to identically vanishing $N_W$ due to the axial symmetry, and we have consistently observed that the total CS number is $N_{\rm CS} \ll 1$ oscillating around zero. This behavior can actually be seen in the first snapshot of Fig.\,\ref{fig:bubble-snapshots}, where a two--bubble collision produces approximately equal and opposite CS number density.

\section{Runaway bubbles}
\label{sec:runaway}

As mentioned in the main text, our simulations apply to electroweak bubbles that collide in the runaway regime. For this case to be physically realized, the vacuum pressure given by the zero-temperature potential needs to overcome the friction from particles gaining mass across the wall in the ballistic regime\,\cite{Bodeker:2009qy}. Moreover, the boost factor at collision, $\gamma_\star$, needs to be smaller than the one corresponding to the terminal velocity imposed by balancing the next-to-leading order pressure from particle splitting, $\mathcal{P}_{\rm NLO} \sim \alpha_w \gamma_w m_W T_n^3$\,\cite{Bodeker:2017cim} (see also\,\cite{Gouttenoire:2021kjv,Azatov:2020ufh,Azatov:2023xem,Baldes:2024wuz} for recent work), where $T_n$ is the nucleation temperature, with $\Delta V \sim v^4$, as typical for a supercooled phase transition. Runaway collisions take place then if $\gamma_w \sim 4 \pi^2 (v/g T_n)^3$ is larger than $\gamma_\star = R_\star/R_c \sim v R_\star$, with $R_\star \simeq (8\pi)^{1/3} \beta^{-1}v_w$. Taking $H\sim v^2/M_{\rm Pl}$ for the Hubble expansion, one obtains $\beta/H \gtrsim 0.1 \,(T_n/{\rm MeV})^3$. This condition may be satisfied for low nucleation temperatures or large values of $\beta/H$, i.e. for fast phase transitions.
For instance, $T_n \sim 10$ MeV requires $\beta/H \sim 10^2$, while
$T_n \in [0.1-1]$ {GeV} leads to $\beta/H \in [10^5 - 10^8] $. The latter (large) values of $\beta/H$ could in fact be realized in scenarios with a seeded electroweak phase transition as explored in Refs.\,\cite{Blasi:2022woz,Agrawal:2023cgp,Blasi:2023rqi}, where $\beta/H$ is effectively replaced by the number of defects in the Hubble volume at the time of bubble nucleation, $\xi$, which can be as large as $\xi \sim 10^8$ for a domain wall network in the friction--dominated regime at the electroweak scale.

\begin{figure}[b!]
\includegraphics[width=8.8cm]
{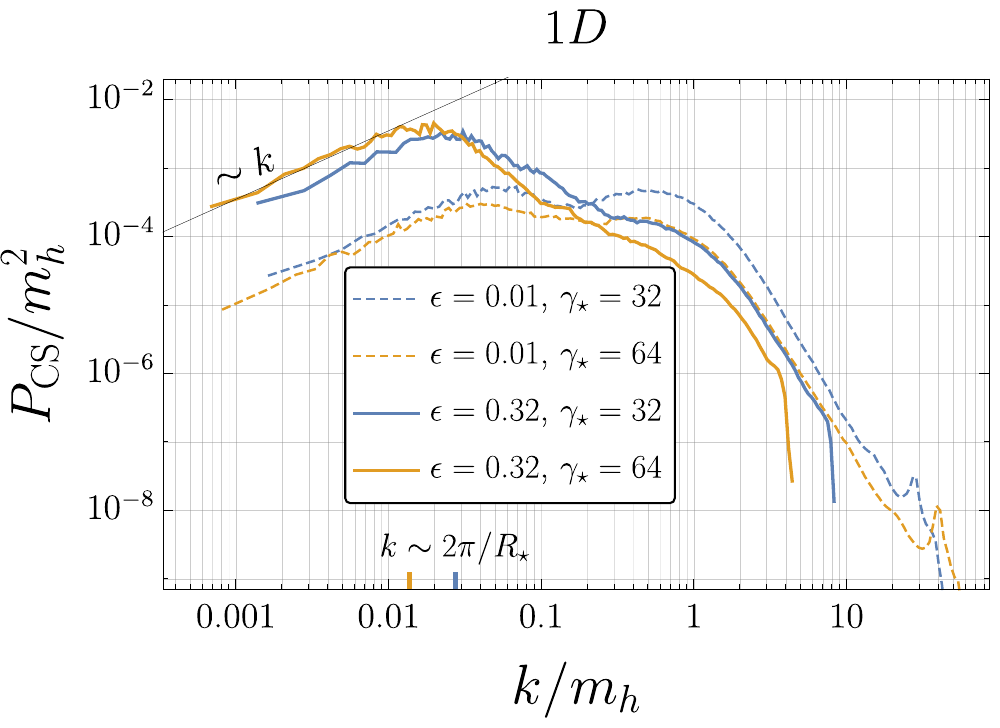}
\hspace{2mm}
\includegraphics[width=8.7cm]{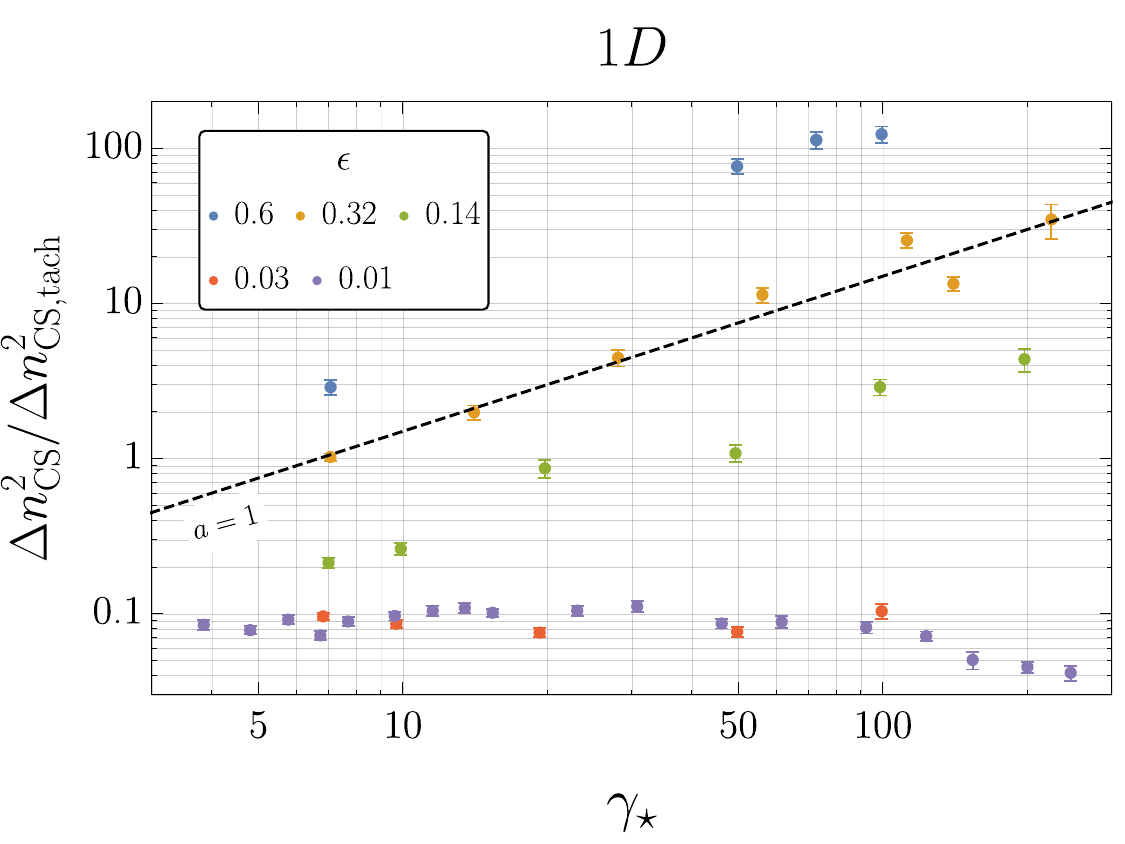}
\vspace{0.2 cm}
\includegraphics[width=8.88cm]{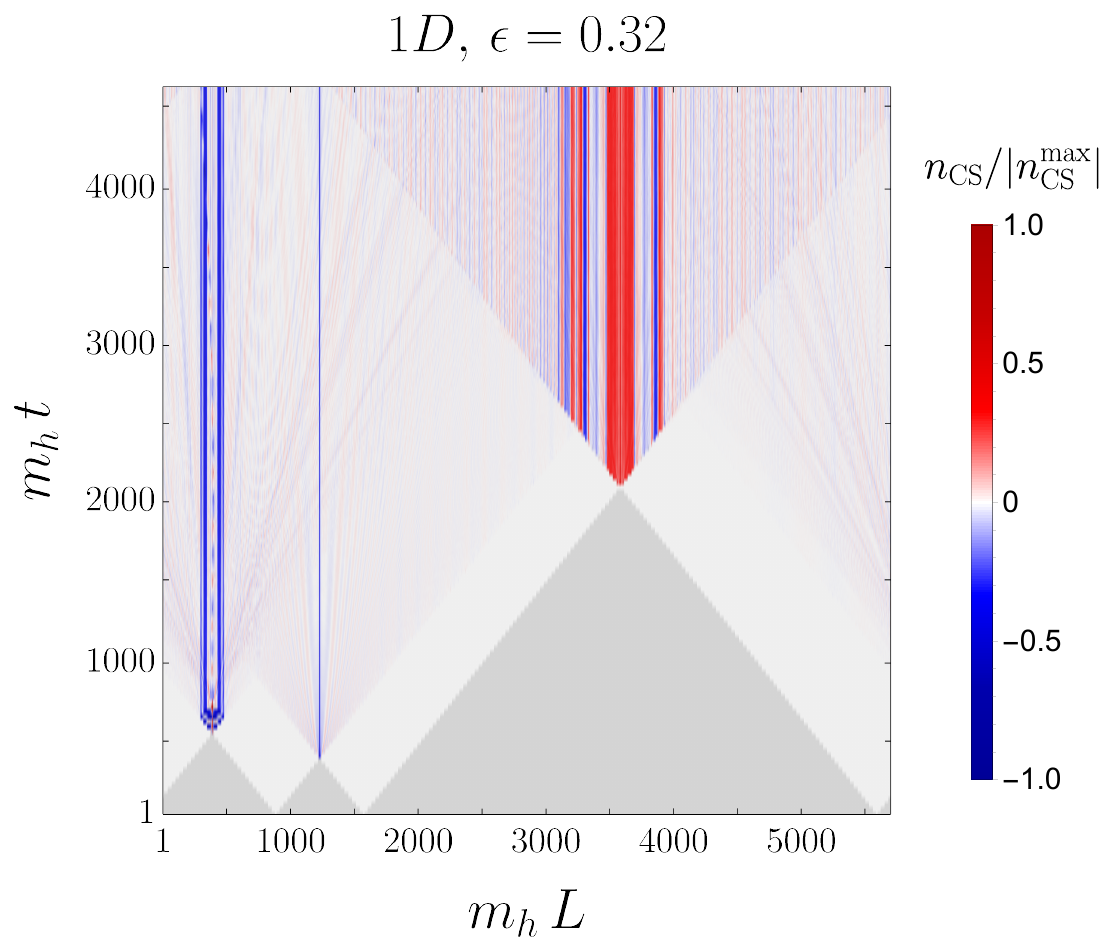} 
\includegraphics[width=8.88cm]{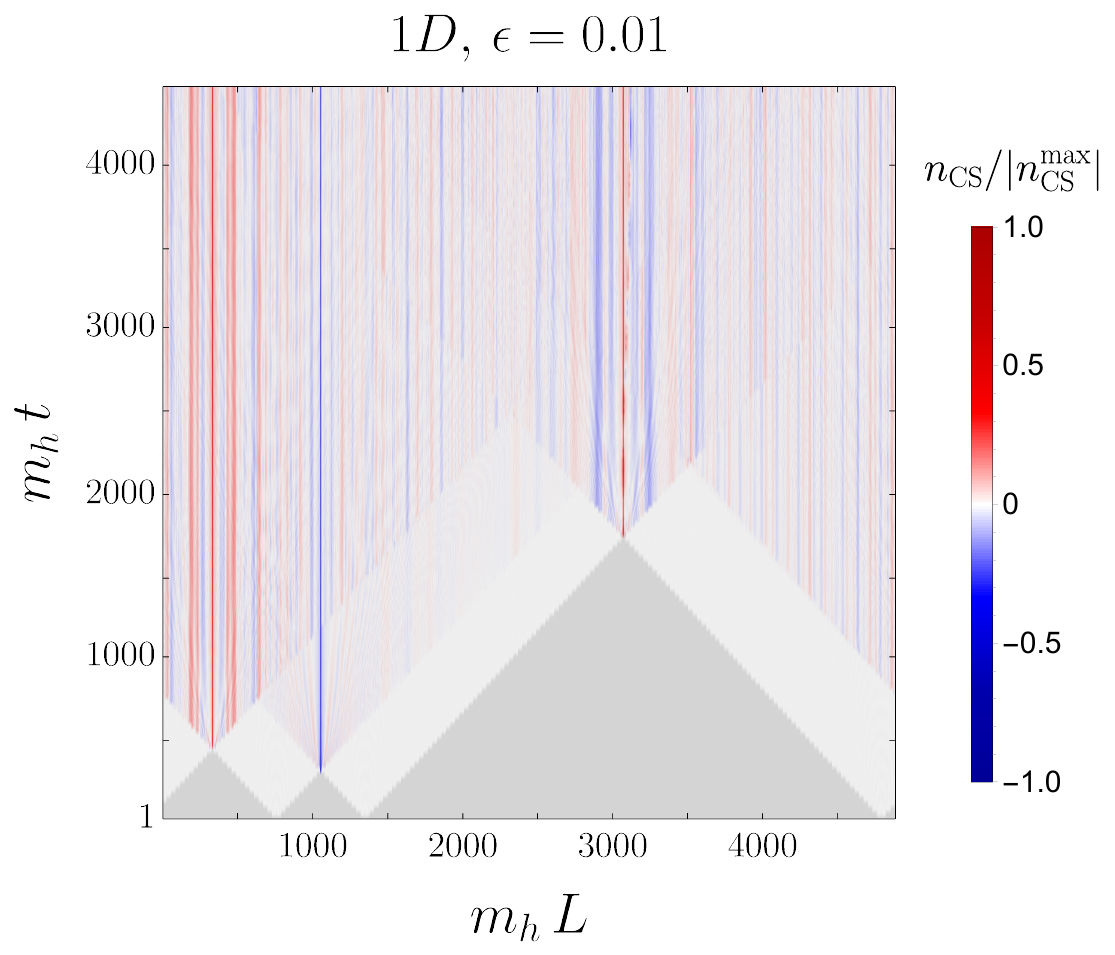}
    \caption{(1+1)D simulations. Top panel, left: The power spectrum of $n_{\rm CS}$, for small and large degeneracy parameter, $\epsilon=\{0.01,0.32\}$. The power spectrum peaks at the scale corresponding to bubble radius at collision $R_\star$, i.e. momentum $2 \pi/R_\star$ (blue and orange ticks on the $x$-axis). For $\epsilon=0.32$, the coefficient of the IR slope at  $k< 2 \pi/R_\star$ (proportional to $n^2_{\text{CS}}$) is higher at larger $\gamma_\star$, indicating an (approximately linear) increase in $n^2_{\text{CS}}$ for increasing  $\gamma_\star$. Top panel, right: CS variance (averaged over 100 realizations) %
    as a function of $\gamma_\star$, for different values of $\epsilon$, normalized by that produced in a tachyonic transition with instantaneous quench, $\Delta n^2_{\rm CS, tach}=0.4 m_h$. 
    Bottom panel: Time-evolution of a 3-bubble collision, with $\epsilon=0.32$ (left), $0.01$ (right). The unbroken phase is represented by dark gray region, while the broken one by light gray. The CS number density is shown in blue and red, normalized to its maximal value attained in the simulation. 
    }
    \label{fig:1D_spectrum}
\end{figure}

\section{Results from (1+1)D simulations of Higgs bubble collisions}
\label{app:results_1D}

To explore large collision boost factors, $\gamma_\star \gtrsim \mathcal{O}(10)$, desirable for extrapolation to the physical point, as well as scalar potentials with $\epsilon \gtrsim 0.1$ that produce large critical bubbles, we performed simulations of runaway bubbles in (1+1)D. These significantly reduce computational cost compared to the (3+1)D simulations presented in the main text, allowing us to extend the simulation box more efficiently. %
For our (1+1)D simulations we have considered the abelian Higgs model with a spontaneously broken $U(1)$ gauge symmetry already adopted in Ref.\,\cite{Garcia-Bellido:1999xos}, as this captures the main topological properties of the SM in (3+1)D. The Higgs potential is modified in order to have a first order phase transition, and we have implemented a scalar potential analogue to the one used in the (3+1)D simulations, as presented in Suppl. Mat.\,\ref{sec:Appendix-Lattice}. 

Our main results are displayed in Fig.\,\ref{fig:1D_spectrum}. %
As shown in the top-left panel, the CS power spectrum exhibits markedly different IR behavior depending on $\epsilon$. For $\epsilon = 0.32$ (solid lines), a peak appears at $k \sim 2\pi/R_\star$, which shifts toward the IR as collisions go from $\gamma_\star = 32$ to $\gamma_\star = 64$, consistently tracking $2\pi/R_\star$ (recall that $R_\star/R_c = \gamma_\star$). %
Since the CS variance is proportional to the amplitude of the IR tail at $k \ll 2\pi/R_\star$, which scales as $\propto k$ by causality in (1+1)D, the CS variance is observed to increase with $\gamma_\star$ (top-right panel of Fig.\,\ref{fig:1D_spectrum}). In contrast, for $\epsilon = 0.01$, the IR power spectrum does not increase with $\gamma_\star$, leading to a constant or decreasing CS variance. As in (3+1)D, a smaller and smaller peak appears at $k \sim m_h$, but it does not contribute to the CS variance.

Despite the clear differences between the (1+1)D and (3+1)D systems, the Higgs potential shape, encoded in $\epsilon$, strongly influences CS production in both cases. In (1+1)D, the variance is significantly larger than that from the tachyonic instability and grows approximately linearly with $\gamma_\star$, as indicated by the $a=1$ line in the top-right panel of Fig.\,\ref{fig:1D_spectrum} (using the same parameterization of Eq.\,\eqref{eq:estimateCS} for the CS variance). Whether such linear growth occurs in (3+1)D remains an open question.

Finally, we comment on the spatial dependence of CS transitions in this lower-dimensional setup, as seen in the bottom panel of Fig.\,\ref{fig:1D_spectrum}. CS number appears to be predominantly produced along the trajectories of collided bubble walls and is not reprocessed afterward. This corresponds to straight lines along the time direction with the same value of the CS number density, $n_{\rm CS}$.  Since the collided walls eventually sweep an $\mathcal{O}(1)$ fraction of the volume, CS production is not purely a surface effect. The (3+1)D analogue of Fig.\,\ref{fig:1D_spectrum}-bottom panel, namely Fig.\,\ref{fig:bubble-snapshots}, shows a less clear indication for the spatial dependence of the CS production (i.e., whether it follows precisely the collided walls), but we expect that similar dynamics may take place in (3+1)D.  Additionally, for $\epsilon = 0.32$ (left panel), regions with sizable CS number are much wider than for $\epsilon = 0.01$ (right panel). For $\epsilon = 0.32$ one can also notice the re-formation of pockets of the false vacuum after the initial collision, as seen for the left-most collision in bottom panel of Fig.\,\ref{fig:1D_spectrum} (left).

\begin{center}
\textbf{\it\Large Supplemental Material
}
\end{center}

\section{Setup and details on the numerical simulation}
\label{sec:Appendix-Lattice}
We only consider the $SU(2)$ gauge group and set $g'=0$. The Higgs field $\phi$ is an $SU(2)$ doublet. 
During the phase transition the Higgs $\phi$ and the $SU(2)$ gauge bosons $W_\mu^a$ follow their classical equations of motion (EoM), derived from the bosonic part of the SM Lagrangian:
\begin{equation}\label{eq:lagrangianSM}
\mathcal{L}= -|D_\mu \phi|^2-\frac12{\rm Tr}[W_{\mu\nu} W^{\mu\nu}]-V(\phi) \, , 
\quad
V(\phi)= \frac{m_h^2}{2v^2} \left(|\phi|^2-\frac{v^2}{2}\right)^2+\frac{1}{\Lambda^2} \left(|\phi|^2-\frac{v^2}{2}\right)^3 +\dots \, .%
\end{equation}
Here, $D_{\mu}\phi=\partial_\mu \phi-ig W_\mu \phi$ is the $SU(2)$-covariant derivative with gauge coupling $g$, $W_\mu=W_\mu^aT_a$ with $T_a=\sigma_a/2$, and $W_{\mu\nu}=\partial_\mu W_{\nu}-\partial_\nu W_{\mu}-ig[W_\mu,W_\nu]$ is the $W$ boson field strength.%
 
As a single-field toy model that induces a  first-order electroweak phase transition (EWPT), we consider a sextic potential -- i.e., we retain only the terms shown in Eq.~\eqref{eq:lagrangianSM}. %
The potential is minimized at the vacuum expectation value $|\phi|^2=v^2/2$, and leads to the canonically normalized Higgs  $h=\sqrt{2}|\phi|$ with mass $m_h$. The mass of the $W$ boson is $m_W=gv/2$. %

The regime of interest is characterized by a low cutoff $\Lambda$, i.e.,
\begin{equation}\label{eq:lambdalim}
\frac{v^4}{m_h^2}<\Lambda^2<\frac{3}{2}\frac{v^4}{m_h^2} \ ,
\end{equation}
where the potential develops an additional local minimum at the origin $h=0$ even at zero temperature, $T=0$. For $\Lambda$ larger than the upper limit in Eq.~\eqref{eq:lambdalim}, %
the origin is a maximum, while for $\Lambda$ smaller than the lower limit, $\sim 484$\,GeV, the origin becomes the global minimum. The potential's shape depends only on the dimensionless cutoff $\Lambda^2m_h^2/v^4$. The potential difference $\Delta V$ between vacua and the degeneracy parameter $\epsilon$ that determines the barrier height with respect to $\Delta V$ are %
\begin{equation}\label{eq:epsLambda}
\Delta V= \frac{m_h^2 v^2}{8} \left(1-1/\ell\right) \, , \qquad \epsilon=\frac{1}{4}\left(1+3/\ell\right) \left(2 -3/\ell^{2}\right)^2 \,  , \quad\qquad  \ell\equiv\frac{\Lambda^2m_h^2}{v^4} \, .
\end{equation}
$\epsilon$ takes value within  $1>\epsilon>0$ for $\Lambda$ in Eq.~\eqref{eq:lambdalim}, with $\epsilon=1$ and $0$ corresponding to degenerate vacua and no barrier, respectively. In Fig.~\ref{fig:Veps} we show $V$ for different $\Lambda$ and related $\epsilon$, as well as $\Lambda^2m_h^2/v^4\to \infty$, i.e. the SM. We will use $\epsilon$ or $\Lambda$ interchangeably according to Eq.~\eqref{eq:epsLambda}. %

The EWPT with the potential in Eq.~\eqref{eq:lagrangianSM} has  been studied in several works, see e.g. Refs.~\cite{Grojean:2004xa,Delaunay:2007wb,deVries:2017ncy}. The finite-temperature corrections to $V$ are non-trivial and phase transition occurs via thermal tunneling and bubble nucleation. In principle, the transition completes only for sufficiently large cutoffs, $\Lambda \gtrsim 550$ GeV. The nucleation temperature in this case is relatively low, $T_n \gtrsim 50$ GeV, and the phase transition strength is $\beta/H\gtrsim 100$~\cite{Grojean:2004xa}, with friction likely playing a significant role on the bubble expansion (see Sec.\,\ref{sec:runaway} in the End Matter). For smaller $\Lambda$, the false vacuum is too deep resulting in a too low nucleation rate, and the Universe inflates. Instead, for $\Lambda\gtrsim 1$~TeV the thermal transition is purely second order. We stress that the potential in Eq.~(\ref{eq:lagrangianSM}) is a technical trick to simplify the simulation. A more realistic modeling of the first-order EWPT that we plan to investigate in the future, will be to include a second scalar field $\sigma$ coupling to the Higgs. A particularly natural way to obtain a strongly supercooled phase transition is to have a nearly conformal potential in the $\sigma$ direction  while the  potential can remain quite standard in the Higgs direction, see e.g. \cite{Bruggisser:2022rdm}. In this case, we can expect qualitatively different dependence of the CS production on the global shape of the scalar potential than the one obtained for the simplified single-scalar-field model.

In the following, we neglect that the transition occurs at a finite temperature and instead analyze bubble nucleation, collision,  gauge boson production and associated effective sphaleron rate assuming the zero-temperature potential in Eq.~\eqref{eq:lagrangianSM}. 
In this regard, let us however mention a possible realization where bubble nucleation is actually compatible with a large potential barrier at $T=0$, corresponding to $\epsilon \sim \mathcal{O}(1)$. While this barrier would typically prevent a fast enough nucleation rate, the presence of topological defects at the time of the phase transition can exponentially enhance the rate in their vicinity, as shown in Refs.\,\cite{Blasi:2022woz,Agrawal:2023cgp,Blasi:2023rqi} for the simplest extension of the SM featuring a singlet scalar field. While the potential barrier around the defects can be effectively smaller and lead to successful nucleation, the dynamics of bubbles at collision will be instead controlled by the actual shape of the potential with $\epsilon \sim \mathcal{O}(1)$, which allows for a larger baryon asymmetry according to the results of our work. This would reconcile efficient bubble nucleation with the optimal type of wall-wall collisions for the production of CS number.

\subsection{Initial conditions and critical bubble}

For transitions with strengths  $\beta/H \gtrsim 1$, the Hubble expansion rate is comparable to or slower than the bubble expansion rate. As a result, the effects of cosmic expansion on the bubble evolution, collision and Chern--Simons number production are expected to be only of order one or smaller. We thus study the EWPT in flat spacetime. To a good approximation the nucleated bubbles are critical Higgs configurations  $h_c(r)$. These are the $O(4)$  solutions of the Euclidean action's equation of motion with potential $V$ in Eq.~\eqref{eq:lagrangianSM}, i.e. 
\begin{equation}
    h''(r) + \frac{3}{r} h'(r) = \partial_{h}V(h/\sqrt{2}) \, ,
\end{equation}
with boundary condition $h'(0)=0$ and $h(r\to\infty)=0$. One has $h_c(0)\simeq v$ so that they interpolate between the true and false vacua. We define their typical radius $R_c$ as $h_c(R_c)=h_c(0)/2$. The left panel in Fig.~\ref{fig:bouncesol} shows the bubble profile for different values of $\epsilon$. The bubble radius $R_c$ is of order $m_h^{-1}$, however as $\epsilon$ tends to $1$ and the vacua at $h=0$ and $h=v$ become degenerate, $R_c$ becomes parametrically larger than $m_h^{-1}$, see right panel of Fig.~\ref{fig:bouncesol}; indeed, $R_c=3 S_1/\Delta V$ where $S_1$ is the bubble tension. This means that lower $\gamma$ factors can be probed for a fixed lattice spacing. The thickness of critical bubble wall, $l_c$, is also set parametrically by $m_h^{-1}$, and  $l_c\sim 3m_h^{-1}$ for the values of $\epsilon<1$ shown in Fig.~\ref{fig:bouncesol}.

\begin{figure}%
\centering
\includegraphics[width=9.2cm]{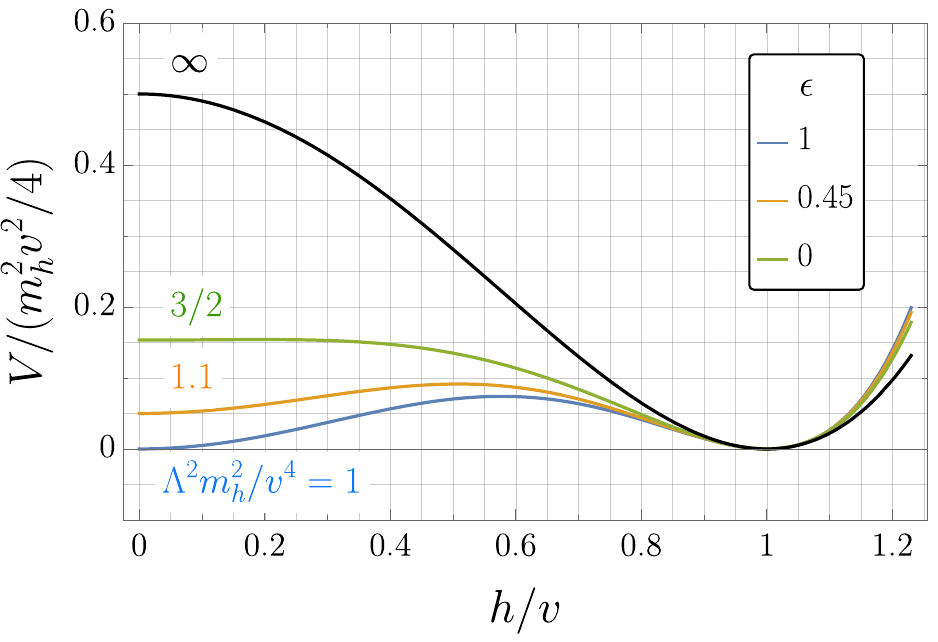}
\caption{The Higgs potential as a function of the cutoff $\Lambda$. The blue, orange and green lines correspond to $\Lambda^2m_h^2/v^4=\{1,1.1,3/2\}$, while the black line to the SM limit $\Lambda^2m_h^2/v^4\to \infty$. The corresponding values $\epsilon$ are also shown. A first-order phase transition at zero temperature occurs for $v^4/m_h^2<\Lambda^2<(3/2)v^4/m_h^2$, and correspondingly $1>\epsilon>0$, where the potential barrier is large or small, respectively.}
\label{fig:Veps}
\end{figure}

Bubbles are expected to nucleate at random spatial positions with random orientations in the Higgs vacuum manifold $SU(2)\simeq S^3$, i.e. 
\begin{equation}\label{eq:phi_b}
\phi_b(r)=\frac{1}{\sqrt{2}}e^{-i \sigma_i \theta_i} (0, h_c(r)) \ ,
\end{equation}
where $\theta_i$ are three random angles, different for each bubble. To a good approximation, all the bubbles are nucleated at the same time $t=0$. The scalar field associated to the initial condition is simply the sum of single bubble configurations $\phi_b$ at rest. The gauge field and its derivatives vanish, $W_\mu=W_{\mu\nu}=0$, consistently with the Gauss constraint (see below).

\subsection{Higgs and gauge fields evolution}
After nucleation, the bubble %
radius increases as $R(t)=\sqrt{t^2+R_c^2}$. The wall velocity is $v_w=\dot{R}$ and its $\gamma$ factor is $\gamma=(1-v_w^2)^{-1/2}$, so that $R=\gamma R_c$ and the wall quickly becomes relativistic. At the same time, the effective wall thickness $l_w$ decreases as $l_w=l_c/\gamma$, and gets smaller than $m_h^{-1}$. As the bubbles collide, the evolution turns nonlinear following the EoM from Eq.~\eqref{eq:lagrangianSM},
\begin{align}\label{eq:EoM1}
&D_\mu D^\mu \phi - \partial_{\phi^*}V(\phi) = 0 \, ,   \\ \label{eq:EoM2}
 & D_\nu W^{\mu\nu} = J^\mu_a T_a \, , \qquad J^\mu_a=2 g {\rm Im}[\phi^\dagger T_a D^\mu\phi] \, .
\end{align}
We solve these numerically on a cubic lattice with length $L$ and $N_x^3=500^3 - 1500^3$ points, and periodic boundary conditions. Upon redefining time and distances as $(t,x)\to m_h^{-1}(t,x)$, and fields as $\phi\to v \phi$ and $W_\mu\to m_h W_{\mu}/g$, Eqs.~\eqref{eq:EoM1} and~\eqref{eq:EoM2}  depend only on $\epsilon$ and the ratio $m_W^2/m_h^2$, fixed to 0.4 based on the SM parameters. %
Although  $m_h/v$ does not enter the EoM, we will use $m_h^2 / v^2 = 1/2$ when showing observables.

The initial conditions have $N_b$ bubbles, so that their radius at collision is approximately $R_\star = \frac{\sqrt{3}}{2}L \big(\frac{3}{ 4 \pi N_b}\big)^{1/3}$ and correspondingly $\gamma_\star=R_\star/R_c$. %
The physical input parameters  are $\epsilon$ and $\gamma_\star$ (or equivalently $R_\star$). %
During the evolution we calculate the volume-average Higgs variance $\langle h^2\rangle\equiv L^{-3}\int d^3x \,h^2(x)$, as well as the average components of the total energy density $\rho_{\rm tot}=T_{00}$, which is conserved:%
\begin{equation}\label{eq:rhotot}
    \rho_{\rm tot}\equiv|\dot{\phi}|^2+|D_i\phi|^2+V(\phi)+\frac{1}{2}\sum_{a,i}{W^a_{0i}}^2+\frac{1}{2}\sum_{a,i<j}{W_{ij}^a}^2 \equiv \rho_K+\rho_G+\rho_V+\rho_E+\rho_B
\end{equation}
where %
$\rho_E={E_i^a}^2/2$ and $\rho_B={B_i^a}^2/2$ are the energy densities of the electric and magnetic fields, $E_i^a=W_{0i}^a$ and $B_i^a=\epsilon_{ijk} W_{jk}^a/2$. We also calculate the Higgs winding number in the volume~$L^3$, defined in the main text.

\begin{figure}%
\centering
\includegraphics[width=8.2cm]{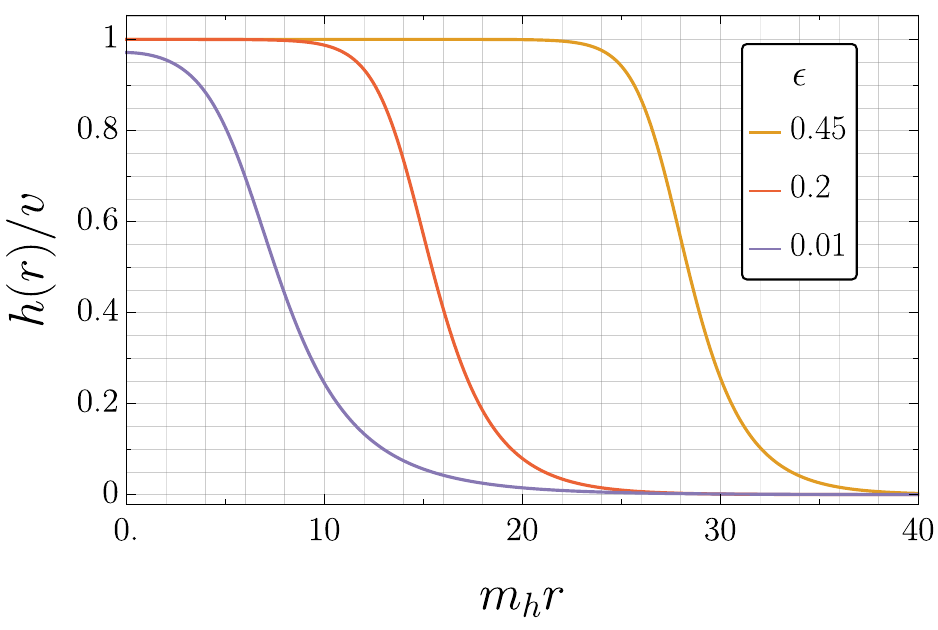} 
\hspace{2mm}
\includegraphics[width=9.2cm]{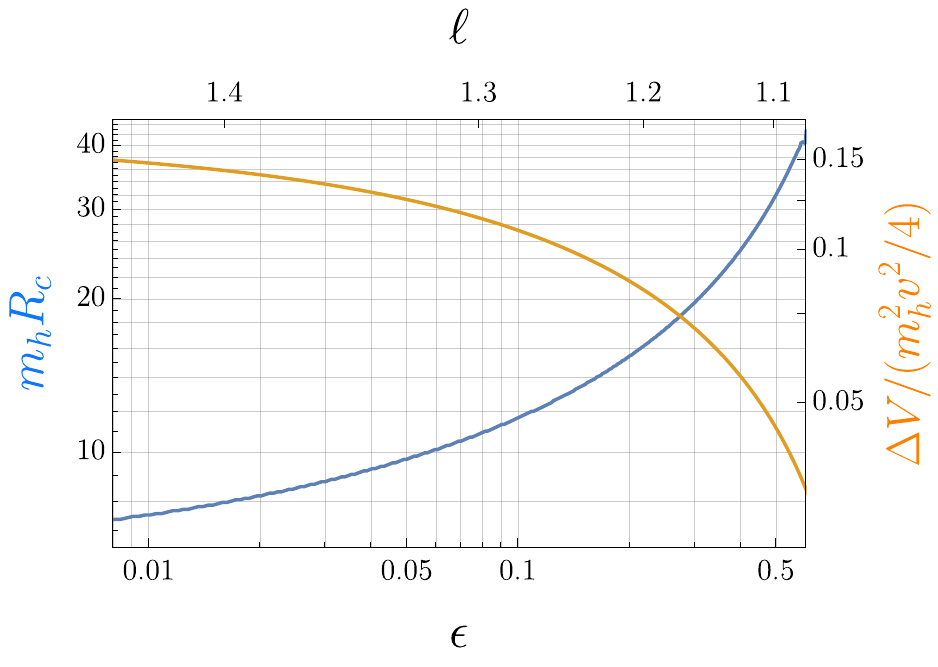}
\caption{Left: Profile of the critical bubble for  $\epsilon=\{0.45,0.2,0.01\}$. Right: Bubble critical radius at nucleation and depth of potential as a function of the degeneracy parameter $\epsilon$ characterizing the potential's shape. }
\label{fig:bouncesol}
\end{figure}

\subsection{Link-plaquette discretization, numerical algorithms and observables}\label{ss:link3d}
Eqs.~\eqref{eq:EoM1} and~\eqref{eq:EoM2} are discretized with standard link-plaquette techniques, which preserve gauge invariance at finite lattice spacing $dx=L/N_x$, and  evolved via a second-order-in-time staggered leapfrog algorithm, with time-step $dt$. For more details see Ref.~\cite{Figueroa:2020rrl}, which we follow.

\vspace{4mm}
\noindent
{\bf Links and plaquettes.} We define the unitary link field matrix $U_\mu(x)\equiv e^{-i g dx^\mu W_\mu(x)}$ going from the space-time point $x+\hat{\mu}$ to $x$, with $dx^0=dt$ and $dx^i=dx$. Although its argument is $x$, $U_\mu$ is to be understood as residing at the midpoint between these two points, namely at $x+\hat{\mu}/2$. Instead, $U_\mu^\dagger$ goes in the opposite direction, i.e. from $x$ to $x+\hat{\mu}$. We denote link field $U_\mu$ at the point $x$ %
shifted by $\hat{\nu}$ as $U_\mu(x+\hat{\nu})\equiv U_{\mu,+\nu}$.  The matrix
\begin{equation} \label{eq:Uij}
U_{\mu\nu}\equiv U_\mu U_{\nu,+\mu}U_{\mu,+\nu}^\dagger U_{\nu}^\dagger=e^{-i g dx^\mu dx^\nu W_{\mu\nu}(1+O(dx^\mu))}
\end{equation}
represents a plaquette centered at a point, with the factors corresponding to its edges ordered counter-clockwise. The second equality can be readily verified by substituting the definition of $U_\mu$. Notice that $U^\dagger_{\mu\nu}=U_{\nu\mu}$. From above, the field strength reads
\beq\label{eq:Wmunu}
W_{\mu\nu}=\frac{i}{2gdx^\mu dx^\nu}(U_{\mu\nu}-U_{\mu\nu}^\dagger)+O(dx^2) \, .
\eeq
For simplicity, we define $U_{-\mu}\equiv U_{\mu,-\mu}^\dagger$, corresponding to the link field entering in $x$ from $x-\hat{\mu}$. Similarly, $\phi_{\pm \mu}\equiv \phi(x\pm \hat{\mu})$ is the scalar field $\phi$ evaluated at $x\pm\hat{\mu}$.

We solve the EoM in the temporal gauge $W_0^a = 0$, so that $U_0 = \mathbf{1}$, and
$U_i = e^{-i g dx W_i}$ are the only relevant link fields. The field's convariant derivative appearing in Eq.~\eqref{eq:EoM1} can be discretized either in the forward or backward directions as
\beq \label{eq:Dphi}
D_\mu^\pm\phi=\pm\frac{1}{dx^{\mu}}\left(U_{\pm \mu}\phi_{\pm \mu}-\phi\right) + O(dx)\, , %
\eeq
 and should be thought as evaluated at $x\pm \hat{\mu}/2$ (as $U_{\pm\mu}$ are). A generic unitary matrix can be expressed in the basis $\bar{\sigma}_\nu \equiv ( \mathbf{1}, i \vec{\sigma} )$ as $U_i = \sum_{\nu=0}^3 c_{i \nu} \bar{\sigma}_\nu$, where the coefficients $c_{i\nu}$ are subject to the normalization  $\sum_{\nu=0}^3 c_{i \nu}^2 = 1$. %

\vspace{4mm}
\noindent
{\bf Evolution equations.} 
 We evolve the 4 real scalar fields $\varphi_i$, defined via the complex doublet $ \phi = (\varphi_1 + i \varphi_2,\; \varphi_3 + i \varphi_4),$ and the 12 components $c_{i \nu}$ of the link variables $U_i$. Along with these, we also evolve the 4+9 components of the field's conjugate momenta $\pi_\phi\equiv \dot{\phi}$ and $\pi_{W,i}^a\equiv W_{0i}^a = \dot{W}_i^a$. The last equality follows from $W_0^a = 0$, so that $D_0W_i^a=\dot{W}_i^a$.  We use a staggered leapfrog evolution scheme, where at each time step the field variables $\phi$ and $U_i$ are evaluated at time $t$, while the  momenta $\pi_\phi$ and $\pi_{W,i}^a$ at time $t + dt/2$.

\vspace{-2mm}
 \begin{itemize}
  \addtolength{\leftskip}{-.8em}
 \item The evolution of $\phi(t-dt)\to\phi(t)$ and $\pi_\phi(t-dt/2)\to\pi_\phi(t+dt/2)$ is obtained expanding $\dot{\phi}(t-dt/2)$ as $(\phi(t)-\phi(t-dt))/dt +O(dt)$, etc.:
\begin{align}\label{eq:phit}
\phi(t)&=\phi(t-dt)+dt \, \pi_\phi(t-dt/2)+O(dt^2)  \, ,   \\ \label{eq:piphit}
\pi_\phi(t+dt/2)&=\pi_\phi(t-dt/2)+ dt \, \ddot{\phi}(t) +O(dt^2) \, ,
\end{align}
$\ddot{\phi}(t)$ in Eq.~\eqref{eq:piphit} is then replaced with the discretized version of Eq.~\eqref{eq:EoM1}:
\beq
\ddot{\phi}(t)=\frac{1}{dx^2}\sum_{i}(U_{i}\phi_{+i}-2\phi+U_{-i}\phi_{-i})-\partial_{\phi^*} V(\phi) \, ,%
\eeq
where right hand side (RHS) is calculated at time $t$. The second covariant derivative $D_i^2\phi$ has been evaluated as $D_i^+D_i^-\phi$. 

\item 
The evolution of $U_i$ (and $\pi_{W,i}^a$) is slightly more complicated. For the former, we evaluate $\dot{U}_i=-igdx \pi_{W,i}U_i$ with $\pi_{W,i}\equiv\pi_{W,i}^aT_a$ at the time $t-dt/2$. In the RHS, we then substitute $\dot{U}_i(t-dt/2)=(U_i(t)-U_i(t-dt))/dt +O(dt)$ and in the left hand side (LHS) we approximate $U_i(t-dt/2)$ with the time average $(U_i(t)+U_i(t-dt))/2$. This leads to the evolution rule:
\begin{align}\label{eq:Uit}
U_i(t)=& \left[1+\frac{igdt dx}{2}\pi_{W,i}(t-dt/2)\right]^{-1}\left[1-\frac{igdt dx}{2}\pi_{W,i}(t-dt/2)\right]U_i(t-dt) + O(dt^2) %
\end{align}
We simplify the product of the square brackets to $[1 - i g dt dx / 2 \,  \pi_{W,i}(t - dt/2)]^2$, since the difference is ${O}(dt^2)$.

\item Similarly to Eq.~\eqref{eq:piphit}, the evolution of $\pi_{W,i}=W_{0i}=\dot{W}_i$ is given by
\beq \label{eq:piWia}
\pi_{W,i}^a(t+dt/2)=\pi_{W,i}^a(t-dt/2)+dt \, \dot{\pi}_{W,i}^a(t) + O(dt^2)
\eeq
where $\dot{\pi}_{W,i}^a(t)$ follows from the gauge field's EoM, Eq.~\eqref{eq:EoM2},  evaluated for $\mu=i$:
\beq \label{eq:dotpiWia}
\dot{\pi}_{W,i}^a=\dot{W}_{0i}^a = (D_j W_{ji})^a + 2g\, {\rm Im}[\phi^\dagger T_a D_i \phi] \, .
\eeq
Specifically, $\dot{\pi}_{W,i}^a$ in Eq.~\eqref{eq:piWia} is substituted with the discretized version of Eq.~\eqref{eq:dotpiWia}:
\beq \label{eq:dotpiWiadisc}
\dot{\pi}_{W,i}^a=  \frac{2i}{g\, dx^3} \, {\rm Tr} \left[ \left( U_{ji} - U_{j,-j}^\dagger U_{ji,-j} U_{j,-j} \right) T_a \right]
+ \frac{2g}{dx} \, {\rm Im} [ \phi^\dagger T_a U_i \phi_{+i} ] \, ,
\eeq
where $U_{ji,-j}$ denotes $U_{ji}$ at $x - \hat{j}$. The first term in Eq.~\eqref{eq:dotpiWiadisc} is the backward difference approximation of the gauge covariant derivative $D_jW_{ji}$. To derive it, we used that $W_{\mu\nu}$ transforms in the adjoint representation, along with Eq.~\eqref{eq:Wmunu}, noting that both terms in the latter contribute equally. The second term in Eq.~\eqref{eq:dotpiWiadisc} originates from $J_a^\mu$ and is derived using $D_i^+ \phi$ from Eq.~\eqref{eq:Dphi}, noting that only the first term contributes since $\phi^\dagger T_a \phi$ is real.

\end{itemize}

A time step consisting of Eqs.~\eqref{eq:phit}+\eqref{eq:Uit}, followed by Eqs.~\eqref{eq:piphit}+\eqref{eq:piWia}, defines a fully algebraic update rule that evolves initial data (given by $\phi$ and $U$ at time $t - dt$, and $\pi_\phi$ and $\pi_{W,i}$ at time $t - dt/2$) forward by a time step $dt$.

Note that this algorithm requires the conjugate momenta to be evaluated half a time step ($dt/2$) ahead of the fields. However, our initial conditions are specified at $t = 0$: the field $\phi(t = 0)$ is given by a superposition of bounce solutions $\phi_b(x)$ as defined in Eq.~\eqref{eq:phi_b}, and the initial momenta and gauge fields vanish, i.e. $\pi_\phi(t = 0) = 0$, $W_i(t = 0) = 0$, and $\pi_{W,i}(t = 0) = 0$. To obtain the values $\pi_\phi(t = dt/2)$ and $\pi_{W,i}(t = dt/2)$ required to initialize the evolution, we apply Eqs.~\eqref{eq:piphit} and~\eqref{eq:piWia} with the RHSs evaluated at $t = 0$ and using $dt \to dt/2$. This corresponds to a single forward step of $O(dt)$.

We work with the dimensionless variables and fields obtained via the rescalings defined below Eq.~\eqref{eq:EoM2}. In this way, only the ratio $m_W^2/m_h^2$ and $\epsilon$ enter the evolution equations Eqs.~\eqref{eq:phit}+\eqref{eq:Uit} and \eqref{eq:piphit}+\eqref{eq:piWia}. We express the evolution equations %
in terms of $\varphi_i$ and $c_{i\nu}$.

\vspace{4mm}
\noindent
{\bf Gauss constraint.} 
The $\mu = 0$ component of the gauge field EoM, which we have not yet used,
\beq
(D_i W_{0i})^a = 2g\, {\rm Im} [ \phi^\dagger T_a \dot{\phi} ],
\eeq
is a constraint equation -- commonly referred to as the Gauss constraint -- since it is first order in time derivatives of $W_i$. This constraint arises from gauge invariance. If it is satisfied by the initial conditions (as it trivially is%
), it is preserved by the evolution provided the EoM hold.

As the simulation progresses, we check the validity of the discretized Gauss constraint:
\beq\label{eq:gaussC}
 \sum_i\frac{\pi_{W,i}(t,x)-U^\dagger_{i,-i}(t)\pi_{W,i}(t,x-\hat{i})U_{i,-i}(t)}{dx} =2 g {\rm Im}[\phi^\dagger(t,x) T_a \pi_{\phi}(t,x)] T_a \,,
\eeq
where we used the fact that $W_{\mu\nu}$ transforms in the adjoint representation and approximated the covariant derivative $D_i W_{0i}$ using a backward finite difference in space. The conjugate momenta $\pi_{\phi}$ and $\pi_{W,i}$ are evaluated at $t+dt/2$ during the evolution (and not $t$), so to evaluate Eq.~\eqref{eq:gaussC} one needs to make a half-step backwards for $\pi$ via Eqs.~\eqref{eq:piphit} and~\eqref{eq:piWia}. We estimate the degree of preservation of the Gauss constraint by calculating the volume-averaged difference and sum of the left- and right-hand side of Eq.~\eqref{eq:gaussC}:
\beq\label{eq:Deltag}
\Delta_g\equiv \frac{\langle|{\rm RHS}-{\rm LHS}|\rangle}{\langle|{\rm RHS}+{\rm LHS}|\rangle} \, .%
\eeq
 Given the gauge invariance of the discretized system, $\Delta_g$ is expected to remain at the level of machine precision -- typically around $10^{-7}$ in single precision and $10^{-14}$ in double precision.

\subsubsection{Observables} \label{ss:observables3d}

\noindent
{\bf Field variance and energies.} The computation of the volume-averaged Higgs field variance, $\langle |\phi|^2 \rangle$, as well as the volume-averaged Higgs kinetic and potential energies, $\langle \rho_K \rangle = \langle |\pi_\phi|^2 \rangle$ and $\langle \rho \rangle = \langle V(\phi) \rangle$, is straightforward. The Higgs covariant gradient energy is obtained as $\langle \rho_G \rangle = dx^{-2} \sum_{i}\langle \left| U_i \phi_{+i} - \phi \right|^2 \rangle$. The electric and magnetic field energies are
$$\langle \rho_E\rangle=\sum_{i,a}\frac12\langle(\pi_{W,i}^a)^2\rangle \, , \quad \langle\rho_B\rangle =\sum_{i,a}\frac12\langle (B_i^a)^2\rangle 
= \sum_{i,j,a}\frac14\langle (W_{jk}^a)^2 \rangle \, ,$$ where $W_{ij}^a = \text{Tr} [ i T_a ( U_{ij} - U_{ji} ) ]/(g\,dx^2)$, see Eq.~\eqref{eq:Wmunu}.

\vspace{4mm}
\noindent
{\bf Higgs winding number.} The Higgs winding number is%
\beq \label{eq:NWnum}
N_W=\frac{1}{24\pi^2}\int d^3x \epsilon_{ijk} \Phi_{ijk} =\frac{1}{8\pi^2}\int d^3x\left(\Phi_{123}-\Phi_{132}\right) \, , \quad \Phi_{ijk}\equiv {\rm Tr}[\Phi^\dagger (\partial_i \Phi)\Phi^\dagger(\partial_j \Phi)\Phi^\dagger(\partial_k \Phi)] \, ,
\eeq
where $\Phi=(i\sigma^2\phi^*,\phi)/|\phi|^2$. To compute $\Phi_{123}$ and $\Phi_{132}$, we first express all the spatial derivatives of $\Phi$ in terms of $\partial_i \phi$, which are then discretized as $\partial_i \phi = (\phi_{+i} - \phi)/dx$, as shown in Eq.~\eqref{eq:Dphi}.

Note that cylindrically symmetric configurations (e.g., along the $z$-axis) satisfy $\Phi_{ijk} = \Phi_{jik}$ and lead to a vanishing $N_W$. This occurs, for example, in the case of two bubbles or three bubbles with centers aligned along a straight line.

\vspace{4mm}
\noindent
{\bf Chern--Simons number.} The topological term ${\rm Tr}[W_{\mu\nu}\tilde{W}^{\mu\nu}]$ that enters $N_{\rm CS}$ %
can be written as the total derivative $\frac{g^2}{16\pi^2} {\rm Tr}[W_{\mu\nu}\tilde{W}^{\mu\nu}]=\partial_\mu j^\mu_{\rm CS}$, with the Chern--Simons current
\beq
j^\mu_{\rm CS}\equiv \frac{g^2}{16\pi^2}\epsilon^{\mu\nu\rho\sigma}\,{\rm Tr}\left[W_\nu W_{\rho\sigma}+\frac23i gW_{\nu}W_{\rho}W_{\sigma} \right]%
\, .
\eeq
The time component of this is the Chern--Simons number density 
\beq\label{n_cs}
n_{\rm CS}\equiv j^0_{\rm CS}=\frac{g^2}{16\pi^2}\epsilon^{ijk}\,{\rm Tr}\left[W_i W_{jk}+\frac23i gW_{i}W_{j}W_{k} \right]%
\, .
\eeq
Using that $\int d^3x\, \partial_i j_{\rm CS}^i = 0$ under periodic boundary conditions (or assuming $j_{\rm CS}^i$ vanishes at spatial infinity), the Chern–Simons number indeed reduces to
\beq
N_{\rm CS}(t)=\int d^3x \, n_{\rm CS}(t,x) \, .
\eeq
We calculate $N_{\rm CS}$ by integrating $n_{\rm CS}$ in space as above. Using $W_i = \frac{i}{2g\,dx} ( U_i - U_i^\dagger )+O(dx)$ and $W_{ij}$ in Eq.~\eqref{eq:Wmunu}, the latter is calculated from the link and plaquette fields as 
\beq
n_{\rm CS} = \frac{g^2}{64\pi^2dx^3} \, \epsilon^{ijk} \ \text{Tr} \left[ (U_i - U_i^\dagger) \left( (U_{jk} - U_{jk}^\dagger) - \frac{1}{3} (U_j - U_j^\dagger)(U_k - U_k^\dagger) \right) \right] +O(dx^3) \, .
\eeq

\vspace{4mm}
\noindent
{\bf Chern-Simons rate.} The Chern-Simons rate measures the number of transitions per unit time and volume between vacua with different Chern–Simons numbers. Formally, it is defined by
\beq
\Gamma_{\rm CS}\equiv\lim_{L,t\to \infty} \frac{\langle N_{\rm CS}^2(t)\rangle - \langle N_{\rm CS}(t)\rangle^2}{L^3t} \, ,
\eeq
where the denominator $L^3 t$ accounts for the fact that the variance of the Chern–Simons number grows linearly with both spatial volume and time in a process that proceeds throughout all space and persists indefinitely. The brackets represent an ensemble average. Because bubble collisions occur over a timescale of order $R_\star$, it is more appropriate to consider the time-dependent sphaleron rate during this transient period:
\beq \label{eq:Gamma_sp_t}
\Gamma_{\rm CS}%
(t)\equiv \lim_{L\to \infty} \frac{d}{dt} \frac{\langle N_{\rm CS}^2(t)\rangle - \langle N_{\rm CS}(t)\rangle^2}{L^3} \equiv \frac{d\Delta n_{\rm CS}^2}{dt} \, .
\eeq
In simulations, the $L \to \infty$ limit in Eq.~\eqref{eq:Gamma_sp_t} is effectively realized by ensuring $R_\star \ll L$. We calculate the ensemble averages in Eq.~\eqref{eq:Gamma_sp_t} by averaging $N_{\rm CS}^2$ and $N_{\rm CS}$ over several different simulations with statistically similar initial conditions%
. The Chern–Simons variance $\Delta n_{\rm CS}^2$ is computed first, followed by evaluation of its time derivative (this smoothed over times $>m_h^{-1})$.

In presenting the results in Fig.~\ref{fig:var_eps02} %
in the main text, the statistical errors over $\Delta n_{\rm CS}^2$ and $\Gamma_{\rm CS}$ are estimated as errors on the variance as follows. First, the Chern--Simons variance $\Delta n_{{\rm CS}, i}^2$ per each run $i=1,\dots,N$ (usually $N=30$) is computed exploiting the white-noise behavior of the Chern--Simons power spectrum for small momenta $k \ll 2\pi/R_\star$ (see discussion in the main text and in the following paragraph), i.e. fluctuations of $N_{\rm CS}$ are uncorrelated for lengths larger than $R_\star$. In this way, the CS variance can also be computed from only one simulation by dividing the whole simulation volume $L^3$ in $n^3$ smaller boxes of volume $(L/n)^3$, such that $L/n \gg R_\star$ (we use $n=4$). The results presented in the main text are obtained using this sub-box division technique. Specifically, the central value of $\Delta n_{\rm CS}^2$ is given by the average $N^{-1}\sum_i\Delta n_{{\rm CS}, i}^2$, while its error %
is estimated as the standard deviation of the values $\Delta n_{{\rm CS}, i}^2$ 
\begin{equation}
    {err}[{ \Delta n_{\rm CS}^2}] = \frac{\sigma_{\Delta n_{{\rm CS}, i}^2}}{\sqrt{N-1}} \, .
\end{equation}

\vspace{4mm}
\noindent
{\bf Power spectra.} The power spectrum $\mathcal{P}_{\varphi}$ of a real field $\varphi$, e.g. the Chern--Simons number density $n_{\rm CS}$%
, is defined in analogy to $P_{\rm CS}$ and satisfies $\langle \varphi^2 \rangle=\int \log k \mathcal{P}_{\varphi}(k)$, as can be immediately seen from its definition. We compute $\mathcal{P}_{\varphi}$ by comparing this last equation to
\beq 
\langle \varphi^2 \rangle= \int \frac{d^3x}{L^3}\varphi^2(\mathbf{x})= 
\int \frac{d^3k}{(2\pi L)^3} | \tilde{\varphi}(\mathbf{k}) |^2 \, ,
\eeq
which leads to
\beq
\mathcal{P}_\phi(|\mathbf{k}|) = \frac{|\mathbf{k}|^3}{(2\pi L)^3} \int d\Omega_k \, | \tilde{\varphi}(\mathbf{k}) |^2 \, , 
\eeq
where $d\Omega_k$ is the solid angle. Here, as before and in the main text, we use the notation $|\mathbf{k}| = k$.%

For a white-noise power spectrum $\mathcal{P}_{\rm CS}=Ck^3/(2\pi^2)$, one can show that $C=\langle N_{\rm CS}^2\rangle/V$, where $V=4\pi R^3/3$ is a spherical volume. Indeed
\beq
\langle N_{\rm CS}^2 \rangle = 
\int_V d^3x \int_V d^3x^\prime \langle n_{\rm CS}(x) n_{\rm CS}(x^\prime) \rangle=\int_V dx \int_V dx^\prime \int \frac{d^3 k}{(2\pi)^3} \int \frac{d^3 k^\prime}{(2\pi)^3} e^{i k x} e^{- i k^\prime x^\prime} \langle \tilde n_{\rm CS}(k) \tilde n_{\rm CS}^*(k^\prime) \rangle \, .
\eeq
(Strictly speaking, the integral over $k$ should be a discrete sum, which is well-approximated by an integral in the large-volume limit.) Using the defintion of $\mathcal{P}_{\rm CS}$ %
one gets
\beq
\langle N_{\rm CS}^2 \rangle  =  \int \frac{d^3k}{4\pi}\frac{\mathcal{P}_{\rm CS}(k)}{k^3}\left(\int_{|\mathbf{x}|<R} d^3x e^{-ikx}\right)^2=\int_0^{\infty}dk\frac{\mathcal{P}_{\rm CS}(k)}{k^3}\frac{16\pi^2[\sin(kR)-kR\cos(kR)]^2}{k^4} \, .
\eeq
Considering a white-noise spectrum peaked at $k_p$, $\mathcal{P}_{\rm CS}=Ck^3/(2\pi^2)$ for $k<k_p$ and $\mathcal{P}_{\rm CS}=0$ otherwise, and using the change of variable $p=kR$, we compute
\beq\label{eq:Ncs2sphere}
\langle N_{\rm CS}^2 \rangle  =   8 C R^3 \int_0^{k_p R} dy \left( \frac{\sin y - y \cos y}{y^2} \right)^2 = \frac{4\pi R^3}{3}C\, .
\eeq
In the LHS we assume that the volume is large enough to resolve the white-noise peak, i.e., $k_p R \gg 1$, in which case the integral evaluates to $\pi/6$. From Eq.~\eqref{eq:Ncs2sphere} it follows $C=\langle N_{\rm CS}^2\rangle/V$. The same result can be obtained by noticing that $N_{\rm CS}$ actually corresponds to the $|k|\rightarrow 0$ modes of $\tilde n_{\rm CS}(k)$.

\subsection{Systematic uncertainties}
\label{app:systunc}

Potential sources of systematic uncertainties in our results are: (1) finite spatial resolution, i.e lattice spacing, set by  $m_hdx$, (2) finite time resolution, set by $m_hdt$, and (3) finite volume effects, set by $R_\star/L$ or equivalently by the number of bubbles $N_b$ in the box. For the main runs we use values of these parameters that are free from these systematics.

\vspace{4mm}
\noindent
{\bf (1) Spatial resolution -- resolving the wall width.} 
The grid spacing must be sufficiently small to resolve the bubble wall. Since the initial wall width is $l_c \sim 3 m_h^{-1}$, this requires $m_h\, dx \lesssim 1$. As the bubble expands, the wall contracts. To resolve the wall width at the time of collision, $l_\star = l_c / \gamma_\star$, the grid must contain few points $l_\star/dx$ per contracted wall. This imposes a more stringent condition:
\begin{equation}
 \frac{l_\star}{dx} = \frac{l_c}{\gamma_\star dx} \gtrsim 1 \,.
\label{eq:wallres_condition}
\end{equation}
We do not calculate the boost factor $\gamma_\star$ explicitly. Instead, we estimate it as
\begin{equation}
    \gamma_\star = \frac{R_\star}{R_c} \simeq \frac{L}{R_c } \frac{\sqrt{3}}{2} \bigg(\frac{3}{4 \pi N_b}\bigg)^{1/3}  \, ,
\end{equation}
where we used that  the mean collision radius is approximately $R_\star = \frac{\sqrt{3}}{2}L \big(\frac{3}{ 4 \pi N_b}\big)^{1/3}$. %

We investigate how our results -- primarily $N_{\rm CS}$ -- depend on the number of points per wall width $l_\star/dx$ by running simulations with $N_b=3$ bubbles with fixed position and Higgs orientations $\theta_i$, corresponding to a fixed $\gamma_\star \simeq 3$. We vary the lattice spacing $dx$ (by changing $N_x$) while keeping all other parameters fixed. Fig.~\ref{fig:wallresolutioN_bis} shows that $l_\star/dx \gtrsim 2$ is sufficient for accurately resolving the  wall%
. We adopt these values for our main runs, except in Fig.~\ref{fig:3D_spectra_big_eps_045} (showing $P_{\rm CS}$ at $\epsilon=0.45$), where achieving a relatively large $\gamma_\star$ required $l_\star/dx \simeq 1$.

\begin{figure}%
\centering
\includegraphics[width=8.5cm]{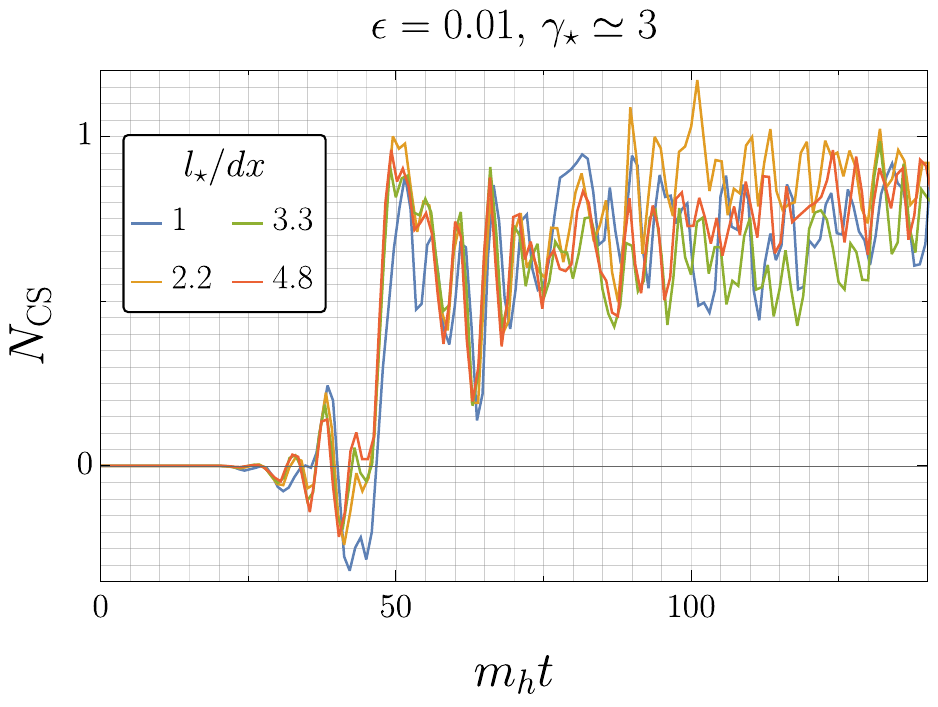}
\caption{Testing the resolution of the bubble wall width at collision $l_\star$: Evolution of total CS number produced from 3-bubble collision for different lattice spacings $dx$.}
\label{fig:wallresolutioN_bis}
\end{figure}

\begin{figure}%
\centering
\subfigure{
\includegraphics[width=8.7cm]{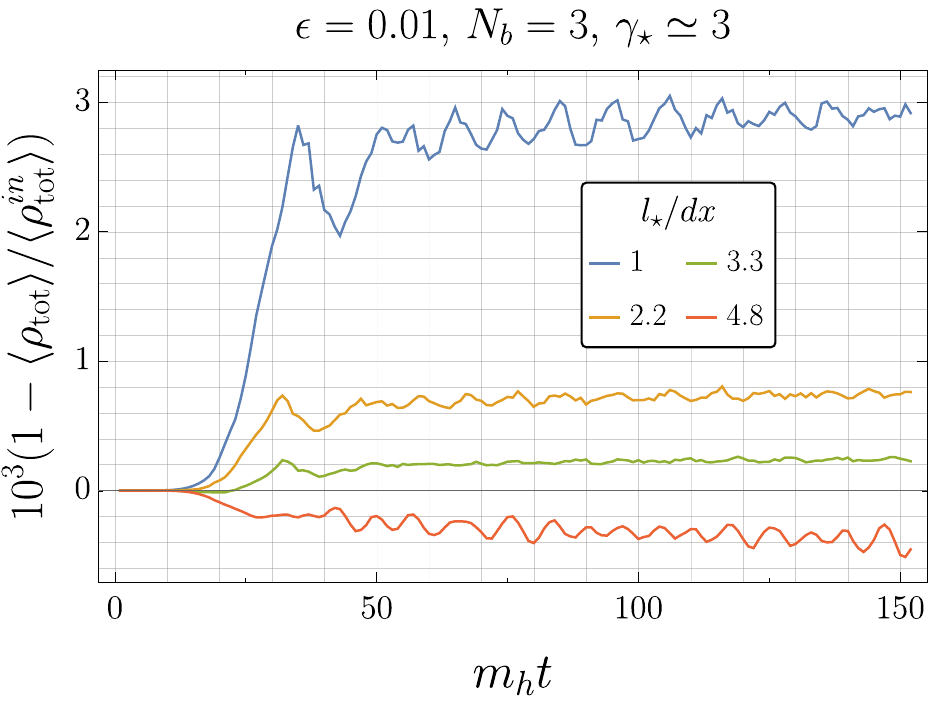}}
\hspace{2mm}
\subfigure{
\includegraphics[width=8.7cm]{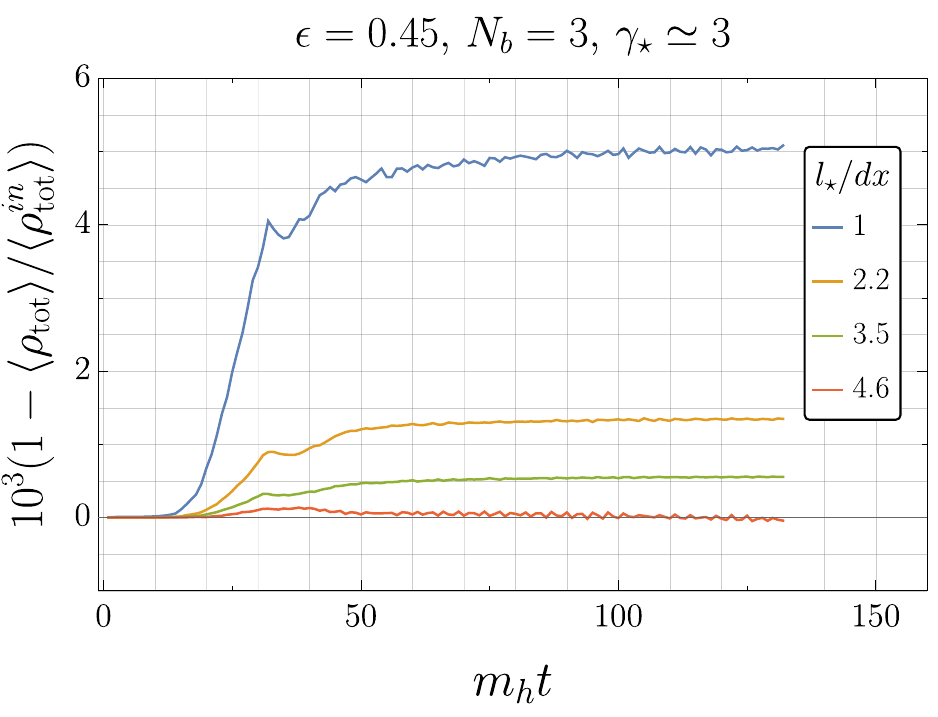}}
\caption{Left panel: energy conservation for a 3-bubble simulation with different lattice spacings (same as Fig.\ref{fig:wallresolutioN_bis}), for fixed $\epsilon=0.01$. Right panel: same as left panel, with $\epsilon=0.45$.}
\label{fig:conservation}
\end{figure}

\begin{figure}%
\centering
\includegraphics[width=8.7cm]{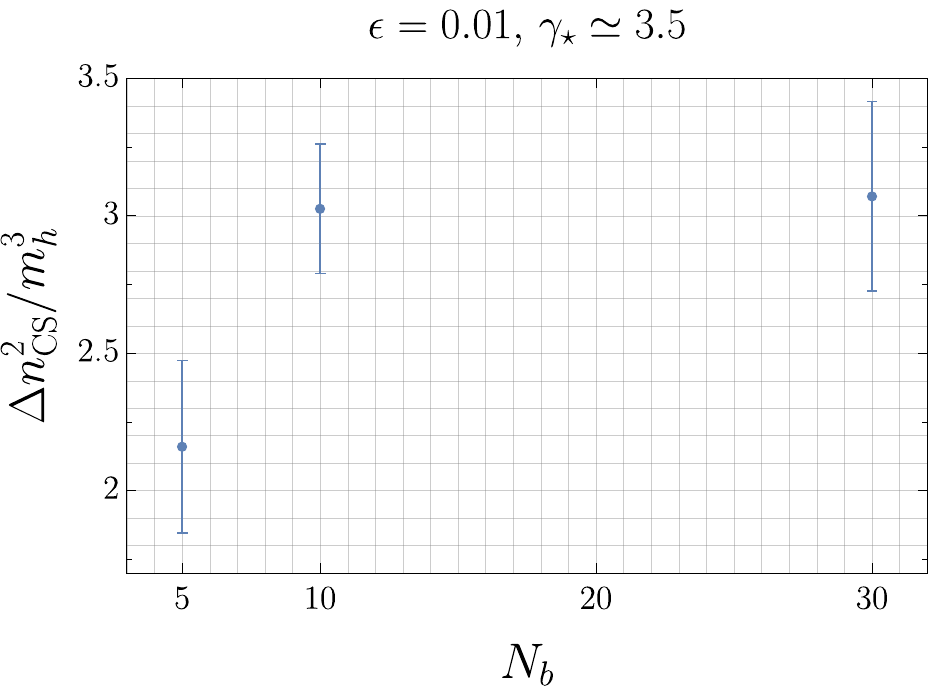}
\caption{Finite volume uncertainties: CS variance for different number of bubbles, i.e. $N_b={5,10,30}$, with fixed $\epsilon=0.01$ and $\gamma_\star \simeq 3.5$. For $N_b\gtrsim 10$, the CS variance values match. The variance is done over 10 simulations.}
\label{fig:bubblenumb}
\end{figure}
\begin{figure}%
\centering
\includegraphics[width=8.7cm]{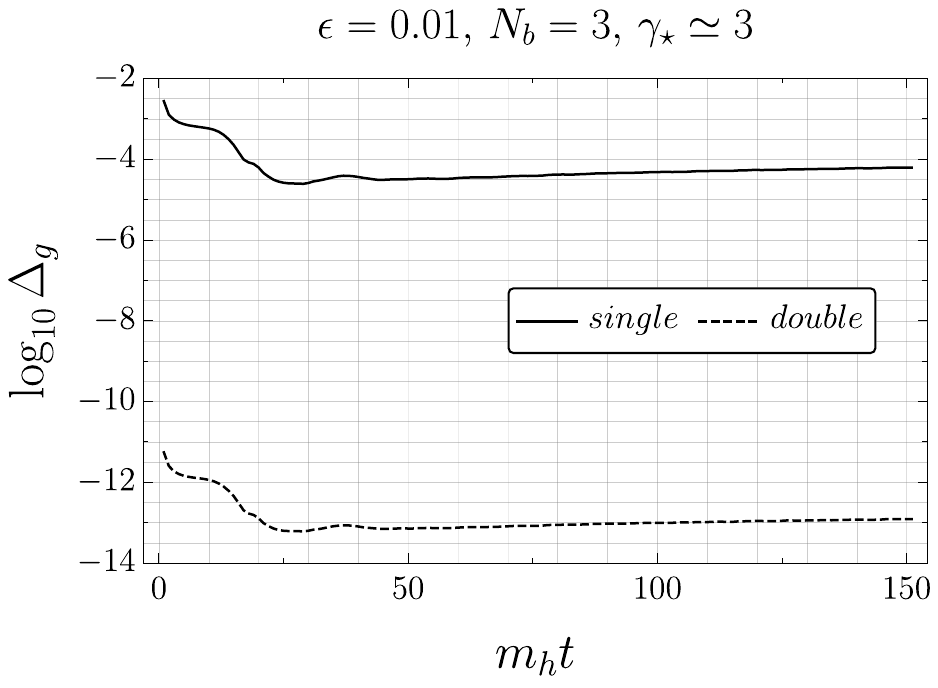}
\caption{ Gauss constraint violation with single- and double-precision variables (same as Fig.\ref{fig:wallresolutioN_bis}).} 
\label{fig:gaussconstr}
\end{figure}

As an additional check of bubble wall resolution, we examine the conservation of the total average energy density $\langle \rho_{\rm tot} \rangle$ (Eq.~\eqref{eq:rhotot})%
. In our run-away bubble scenario, the majority of the energy is carried by the kinetic energy of the bubble walls at collision. Total energy conservation is thus a robust indicator for the validity of the overall evolution. In Fig.~\ref{fig:conservation}, we show the variation of the total energy of 3-bubble simulations, for benchmarks $\epsilon=\{0.01, 0.45\}$. $\langle \rho_{\rm tot} \rangle$ is conserved better than per-mille level, with the largest loss occurring at the time of the first bubble collision. If the number of points per wall, $l_\star/dx$, is less than 2, $\langle \rho_{\rm tot} \rangle$ is however conserved only at the per-mille level.

\vspace{4mm}
\noindent
{\bf (2) Time resolution.} In each simulation, we fixed the time-step as $dt/dx=0.3$. The errors in both $\langle \rho_{\rm tot} \rangle$ and $N_{\rm CS}$ from not using a smaller $dt$ are negligible compared to those arising from finite lattice spacing.

\vspace{4mm}
\noindent
{\bf (3) Finite volume.} %
The Chern–Simons number $N_{\rm CS}$ and the Chern-Simons rate (Eq.~\eqref{eq:Gamma_sp_t}) may be impacted by finite volume effects when the box size $L$ is not significantly larger than the largest characteristic scale of field fluctuations, set by the bubble radius at collision $R_\star$. In particular, periodic boundary conditions and a limited number of bubbles can introduce systematic biases in the generation of $N_{\rm CS}$. Since two bubbles alone do not produce $N_{\rm CS}$, repeated collisions of the same pair -- enabled by periodic boundaries -- can  suppress $N_{\rm CS}$ production. To avoid these effects, we ensure that the ratio $R_\star/L \simeq 0.5\,N_b^{-1/3}$ is sufficiently small -- that is, we use a large enough number of bubbles, $N_b$. Fig.~\ref{fig:bubblenumb} shows that $N_{\rm CS}$ is statistically indistinguishable for $N_b = 10$ and $N_b = 30$, suggesting that $R_\star/L \lesssim 1/4$ is already close to the infinite volume limit. In our main simulations, we use $N_b = 10$.

\vspace{3mm}
Finally, Fig.~\ref{fig:gaussconstr} shows the Gauss constraint violation $\Delta_g$ (Eq.~\eqref{eq:Deltag}), which is a strong check of the accuracy of the gauge field evolution. We find that $\Delta_g$ is conserved at the $10^{-4}$ and $10^{-13}$ levels when using single- and double-precision variables, respectively. For our main simulations, we use single precision, as the associated truncation errors are negligible compared to the systematic uncertainties arising from finite lattice spacing effects.

\section{$U(1)$-gauge field in (1+1)$D$}
\label{app:U(1)}

The 3+1-dimensional $SU(2)$ simulations are constrained to relatively small grid sizes (up to $N_x^3 \sim 1500^3$), limiting the accessible Lorentz factors at collision to $\gamma_\star \lesssim 10$. To gain insight into the behavior of the Chern--Simons rate at higher $\gamma$, we instead study a simpler 1+1-dimensional $U(1)$ model, which still supports texture configurations%
. This lower-dimensional setup allows for significantly lattice sizes up to $N_x\sim 10^7$ points, enabling the exploration of regimes with $\gamma_\star \sim 200$. This setup has been considered in the original work on cold baryogenesis~\,\cite{Garcia-Bellido:1999xos}.

We consider a $U(1)$ classical gauge theory in 1D (spatial dimension) with gauge field $A_\mu$, $\mu=0,1$, and a complex scalar $\psi$ in the fundamental representation. The Lagrangian is
\begin{equation}\label{eq:L1D}
\mathcal{L}=-|D_\mu\psi|^2-\frac14F_{\mu\nu}F^{\mu\nu}-V(\psi) \, , \qquad D_\mu\psi=\partial_\mu\psi-ig A_\mu\psi    \, ,
\end{equation}
with field strength $F_{\mu\nu}=\partial_\mu A_\nu-\partial_\nu A_\mu$ and $V$ is the same potential as in Eq.~\eqref{eq:lagrangianSM}. In 1D, the mass dimensions are $[\psi]=[A_\mu]=[v]=0$, $[g]=[m_h]=1$ and $[\Lambda]=-1$, while $\ell$ in Eq.~\eqref{eq:epsLambda} is still dimensionless. The potential difference between vacua $\Delta V$ and the barrier parameter $\epsilon$ are still given by Eq.~\eqref{eq:epsLambda}. Around the true vacuum, the physical Higgs is $h\equiv\sqrt{2}|\psi|$ and has a mass $m_h$, while the gauge boson mass is $m_A = g v$; note the factor of 2 difference compared to the non-Abelian case. The vacuum manifold is $M=U(1)$. Thus, the theory admits texture configurations labeled by the maps $M=U(1)\to S^1=\mathbb{R}\cup \{\infty\}$, where $\psi$ is always in the vacuum. As one-dimensional bubbles collide, we may therefore expect analogous dynamics in the generation of winding number and Chern–Simons number. The equations of motion (EoM) are
\begin{equation}\label{eq:EoM1d}
D_\mu D^{\mu}\psi=\partial_{\psi^*} V(\psi) \, , \qquad \partial_\nu F^{\mu\nu}=2g {\rm Im}[\psi^*D^\mu\psi]\, .
\end{equation}
Upon redefining  $(t,x)\to m_h^{-1}(t,x)$, and  $\psi\to v \psi$ and $A_\mu\to m_h A_{\mu}/g$, %
they still depend only on $\epsilon$ and the ratio $m_A^2/m_h^2=(gv/m_h)^2$, fixed to 0.4. We solve them %
numerically in a periodic 1D  grid with length $L$ and $N_x\lesssim10^6-10^7$ points. 

\vspace{4mm}
\noindent
{\bf Bubbles in 1D and initial conditions.} Higgs bubbles are segments where the field $h$ remains near the true vacuum $h = v$ over a distance $2R$, with $R$ the bubble radius, and is otherwise at $h = 0$. The critical bubbles nucleated are the $O(2)$-symmetric bounce solutions of the Euclidean EoM, $h_c(x)$,
\begin{equation}\label{eq:bounce1d}
    h''(r)+ \frac{1}{r} h'(r) = {\partial_{h} V(h/\sqrt{2})} \, ,
\end{equation}
with boundary conditions $h'(0)=0$ and $h(x\to\infty)=0$. We define their critical radius $R_c$ and wall thickness $l_c$ in analogy to the three-dimensional case. Both are set parametrically by $m_h^{-1}$ and $l_c\sim 3m_h^{-1}$ for the values of $\epsilon$ we studied. During the expansion, the bubble radius follows $R(t)^2 -t ^2 = R_c^2$. Similarly to 3D, the wall velocity is $v_w=\dot{R}=\sqrt{1-R_c^2/R^2}$ and the boost factor is $\gamma=R/R_c$.

Note that the bounce solution in Eq.~\eqref{eq:bounce1d} only sets $|\psi|$ and not the phase of $\psi$. As initial conditions, we use a superposition of $N_b = 10$ critical bubbles $\psi_c(x) = e^{i\theta} h_c(x)/\sqrt{2}$, each placed at a random position in the domain $L$ and assigned a random phase $\theta$, independent for each bubble. In addition, we set $A_{\mu}=F_{\mu\nu}=0$.

\vspace{4mm}
\noindent
{\bf Winding and Chern--Simons numbers.} The winding and Chern--Simons numbers are
\begin{equation}\label{eq:NcsNw1d}
N_W=\frac{1}{2\pi}\int d x \,\partial_x {\rm Arg}[\psi] \ , \qquad N_{\rm CS}=\frac{g}{2\pi}\int dx A_1  ,
\end{equation}
The Chern--Simons number density is simply $n_{\rm CS}=g A_1/(2\pi)$. 
As mentioned, the theory in Eq.~\eqref{eq:L1D} supports texture configurations with a nontrivial $N_W$, e.g. $\psi(x)=e^{i\chi(x)}$, with $\chi(x)=\pi\tanh(x/r_t)$ has $N_W=1$. The Chern--Simons variance and rate are
\beq
\Delta n^2_{\rm CS}\equiv\frac{\langle N_{\rm CS}^2\rangle-\langle N_{\rm CS}\rangle^2}{L} \, , \qquad \Gamma_{\rm CS}=\frac{d\Delta n^2_{\rm CS}}{dt} \, .
\eeq
Similar to 3D, we also compute the Chern--Simons power spectrum $\mathcal{P}_{\rm CS}$ defined by
\beq \label{eq:Pspec1d}
\langle n_{\rm CS}^*(k)n_{\rm CS}(k')\rangle=\frac{(2\pi)^2}{k} \delta(k-k')\mathcal{P}_{\rm CS}(k) \ .
\eeq
As in 3D, for $k < 2\pi/R_\star$, the spectrum takes a white-noise form,
$\mathcal{P}_{\rm CS}(k) = Ck/\pi$, where, as shown by a calculation similar to that in Suppl. Mat.~\ref{ss:observables3d}, the coefficient is given by $C = \Delta n_{\rm CS}^2$.

\subsection{Discretization and numerical evolution}
We solve Eq.~\eqref{eq:EoM1d} with the same link-plaquette techniques and staggered Leapfrog algorithm as the 3D case in Suppl. Mat.~\ref{ss:link3d}. Here we highlight only the  differences. Although the spatial index is $i = 1$, for generality we express the formulas for an arbitrary number of spatial dimensions. By running large grids in (1+1)D, we can use a particularly fine lattice spacing $dx$, ensuring that the number of points per wall width at collision satisfies $l_\star/dx \gtrsim 8$. As discussed below, this keeps the main quantities close to the continuum limit.

\vspace{4mm}
\noindent
{\bf Links and plaquettes.} The link field $V_\mu\equiv e^{-i g dx^\mu A_\mu(x)}$ that connects $x+\hat{\mu}$ to $x$ is now a phase, and $V_\mu^*$ connects the two points in the opposite direction. The plaquette $V_{\mu\nu}$ is
\beq \label{eq:Vij}
V_{\mu\nu}\equiv V_\mu V_{\nu,+\mu}V_{\mu,+\nu}^*V_{\nu}^*=e^{-i g dx^\mu dx^\nu F_{\mu\nu} (1+O(dx))} \, , \quad F_{\mu\nu}=\frac{i(V_{\mu\nu}-V_{\mu\nu}^*)}{2gdx^\mu dx^\nu}+O(dx^2) \, .
\eeq
Notice that $V^*_{\mu\nu}=V_{\nu\mu}$. We define $V_{-\mu}\equiv V_{\mu,-\mu}^*$, and the forward and backward covariant derivatives are as before 
$D_\mu^\pm\phi=\pm\left(V_{\pm \mu}\phi_{\pm \mu}-\phi\right)/dx^{\mu} +O(dx)\, .$

\vspace{4mm}
\noindent
{\bf Evolution equations.} We use the gauge $A_0=0$, leading to $V_0=1$ and  $V_i=e^{-i g dx A_i}$. We evolve the real and imaginary parts of $\psi$ and of $V_i$, as well as of their conjugate momenta $\pi_\psi\equiv \dot{\psi}$ and $\pi_{A,i}\equiv F_{0i}=\dot{A}_i$. The evolution of $\psi$ is as in Eq.~\eqref{eq:phit}:
\begin{align}\label{eq:psit}
\psi(t)&=\psi(t-dt)+dt \, \pi_\psi(t-dt/2)+O(dt^2)  \, ,   \\ \label{eq:pipsit}
\pi_\psi(t+dt/2)&=\pi_\psi(t-dt/2)+ dt \, \ddot{\psi}(t) +O(dt^2) \, ,
\end{align}
where, from the scalar field EoM in Eq.~\eqref{eq:EoM1d},
\beq
\ddot{\psi}(t)=\frac{1}{dx^2}\sum_{i}(V_{i}\psi_{+i}-2\psi+V_{-i}\psi_{-i})-\partial_{\psi^*} V(\psi) \, .%
\eeq
As in Eq.~\eqref{eq:Uit}, from $\dot{V}_i=-igdx \pi_{A,i}V_i$, it follows the evolution of $V_i$:
\beq
V_i(t)= V_i(t-dt)\left[\frac{1-\frac{igdt dx}{2}\pi_{A,i}(t-dt/2)}{1+\frac{igdt dx}{2}\pi_{A,i}(t-dt/2)}\right]\, + O(dt^2) \, .
\eeq
Similar to Eq.~\eqref{eq:piWia}, for the evolution of $\pi_{A,i}$ we have
$$ \pi_{i,A}(t+dt/2)=\pi_{i,A}(t-dt/2)+ dt \dot{F}_{0i}(t) +O(dt^2) \, , $$ 
where $\dot{F}_{0i}=\partial_jF_{ji}+2g{\rm Tr}[\psi^*\partial_j\psi]$ follows from the EoM in 
Eq.~\eqref{eq:EoM1d}. Its discretized version is
\beq
\dot{F}_{0i}=\frac{i}{g \,dx^3}\sum_j(V_{ji}-V_{ji,-j}) +\frac{2g }{dx} {\rm Im}[\psi^*V_i\psi_{+i}] \, . 
\eeq

\vspace{2mm}
\noindent
{\bf Gauss constraint.} The Gauss constraint $\partial_i F_{0i}=\partial_i\pi_{A,i}=2 g {\rm Im}[\psi^*\dot{\psi}]$ is discretized as:
\beq
 \frac{\pi_{A,i}(t,x)-\pi_{A,i}(t,x-\hat{i})}{dx} =2 g {\rm Im}[\psi^*(t,x)\pi_\psi(t,x)] \,,
\eeq
where as in 3D we make a step backwards for $\pi_\psi$ and $\pi_{A,i}$. $\Delta_g$ in Eq.~\eqref{eq:gaussC} is again used to evaluate the preservation of the Gauss constraint. We use double-precision variables and observe conservation of $\Delta_g$ at the $10^{-10}$ level.

\vspace{4mm}
\noindent 
{\bf Observables.} The average energy is $\langle\rho\rangle=\int \frac{d^3x}{V} \rho$ and
\beq
\rho=|\dot{\psi}|^2+\sum_i|D_i\psi|^2+V(\psi)+\frac12\sum_i(E_i^2+B_i^2)
\eeq
with $E_i=F_{0i}=\pi_{A,i}$, $B_i=\frac12\epsilon^{ijk}F_{jk}$. Note that $B_i=0$ in 1D and $|D_i\psi|^2=dx^{-2}|V_i\psi_{+i}-\psi|^2$. We compute the 1D winding and Chern--Simons numbers of Eq.~\eqref{eq:NcsNw1d} as 
\beq
N_W= \sum_n ({\rm Arg}[\psi(x_n+1)]-{\rm Arg}[\psi(x_n)]) \, , \quad N_{\rm CS}=-\frac{1}{2\pi}\sum_n {\rm Arg}[V_1(x_n)] \, ,
\eeq
Here, $x_n = n\, dx$, and the sums run over all grid points. Note that $\mathrm{Arg}[V_1] = \tan^{-1}(\mathrm{Im}[V_1]/\mathrm{Re}[V_1]) \simeq \mathrm{Im}[V_1]$ to a good approximation. 
With our choice of $dx$, the total energy is conserved to the $10^{-6}$ level.

\newpage 
\end{widetext}

\bibliography{biblio_PRL}

@article{Kuzmin:1985mm,
    author = "Kuzmin, V. A. and Rubakov, V. A. and Shaposhnikov, M. E.",
    title = "{On the Anomalous Electroweak Baryon Number Nonconservation in the Early Universe}",
    reportNumber = "IC/85/8",
    doi = "10.1016/0370-2693(85)91028-7",
    journal = "Phys. Lett. B",
    volume = "155",
    pages = "36",
    year = "1985"
}

@article{Fukugita:1986hr,
    author = "Fukugita, M. and Yanagida, T.",
    title = "{Baryogenesis Without Grand Unification}",
    reportNumber = "RIFP-641",
    doi = "10.1016/0370-2693(86)91126-3",
    journal = "Phys. Lett. B",
    volume = "174",
    pages = "45--47",
    year = "1986"
}

@article{Davidson:2008bu,
    author = "Davidson, Sacha and Nardi, Enrico and Nir, Yosef",
    title = "{Leptogenesis}",
    eprint = "0802.2962",
    archivePrefix = "arXiv",
    primaryClass = "hep-ph",
    doi = "10.1016/j.physrep.2008.06.002",
    journal = "Phys. Rept.",
    volume = "466",
    pages = "105--177",
    year = "2008"
}

@article{Garcia-Bellido:1999xos,
    author = "Garcia-Bellido, Juan and Grigoriev, Dmitri Yu. and Kusenko, Alexander and Shaposhnikov, Mikhail E.",
    title = "{Nonequilibrium electroweak baryogenesis from preheating after inflation}",
    eprint = "hep-ph/9902449",
    archivePrefix = "arXiv",
    reportNumber = "IMPERIAL-TP-98-99-39, UCLA-99-TEP-7, UNIL-IPT-99-1",
    doi = "10.1103/PhysRevD.60.123504",
    journal = "Phys. Rev. D",
    volume = "60",
    pages = "123504",
    year = "1999"
}

@article{Krauss:1999ng,
    author = "Krauss, Lawrence M. and Trodden, Mark",
    title = "{Baryogenesis below the electroweak scale}",
    eprint = "hep-ph/9902420",
    archivePrefix = "arXiv",
    reportNumber = "CWRU-P11-99",
    doi = "10.1103/PhysRevLett.83.1502",
    journal = "Phys. Rev. Lett.",
    volume = "83",
    pages = "1502--1505",
    year = "1999"
}

@article{Cornwall:2000eu,
    author = "Cornwall, John M. and Kusenko, Alexander",
    title = "{Baryon number nonconservation and phase transitions at preheating}",
    eprint = "hep-ph/0001058",
    archivePrefix = "arXiv",
    reportNumber = "UCLA-00-TEP-02",
    doi = "10.1103/PhysRevD.61.103510",
    journal = "Phys. Rev. D",
    volume = "61",
    pages = "103510",
    year = "2000"
}

@article{Copeland:2001qw,
    author = "Copeland, Edmund J. and Lyth, David and Rajantie, Arttu and Trodden, Mark",
    title = "{Hybrid inflation and baryogenesis at the TeV scale}",
    eprint = "hep-ph/0103231",
    archivePrefix = "arXiv",
    reportNumber = "SUSX-TH-01-014, DAMTP-2001-18, SU-GP-01-2-3",
    doi = "10.1103/PhysRevD.64.043506",
    journal = "Phys. Rev. D",
    volume = "64",
    pages = "043506",
    year = "2001"
}

@article{Smit:2002yg,
    author = "Smit, Jan and Tranberg, Anders",
    title = "{Chern-Simons number asymmetry from CP violation at electroweak tachyonic preheating}",
    eprint = "hep-ph/0211243",
    archivePrefix = "arXiv",
    reportNumber = "ITFA-2002-51",
    doi = "10.1088/1126-6708/2002/12/020",
    journal = "JHEP",
    volume = "12",
    pages = "020",
    year = "2002"
}

@article{Garcia-Bellido:2003wva,
    author = "Garcia-Bellido, Juan and Garcia-Perez, Margarita and Gonzalez-Arroyo, Antonio",
    title = "{Chern-Simons production during preheating in hybrid inflation models}",
    eprint = "hep-ph/0304285",
    archivePrefix = "arXiv",
    reportNumber = "FT-UAM-03-07, IFT-UAM-CSIC-03-13",
    doi = "10.1103/PhysRevD.69.023504",
    journal = "Phys. Rev. D",
    volume = "69",
    pages = "023504",
    year = "2004"
}

@article{Tranberg:2003gi,
    author = "Tranberg, Anders and Smit, Jan",
    title = "{Baryon asymmetry from electroweak tachyonic preheating}",
    eprint = "hep-ph/0310342",
    archivePrefix = "arXiv",
    reportNumber = "ITFA-2003-49",
    doi = "10.1088/1126-6708/2003/11/016",
    journal = "JHEP",
    volume = "11",
    pages = "016",
    year = "2003"
}

@article{vanderMeulen:2005sp,
    author = "van der Meulen, Meindert and Sexty, Denes and Smit, Jan and Tranberg, Anders",
    title = "{Chern-Simons and winding number in a tachyonic electroweak transition}",
    eprint = "hep-ph/0511080",
    archivePrefix = "arXiv",
    reportNumber = "ITFA-2005-46",
    doi = "10.1088/1126-6708/2006/02/029",
    journal = "JHEP",
    volume = "02",
    pages = "029",
    year = "2006"
}

@article{Turok:1990zg,
    author = "Turok, Neil and Zadrozny, John",
    title = "{Electroweak baryogenesis in the two doublet model}",
    reportNumber = "PUPT-90-1216",
    doi = "10.1016/0550-3213(91)90356-3",
    journal = "Nucl. Phys. B",
    volume = "358",
    pages = "471--493",
    year = "1991"
}

@article{Tranberg:2006ip,
    author = "Tranberg, Anders and Smit, Jan",
    title = "{Simulations of cold electroweak baryogenesis: Dependence on Higgs mass and strength of CP-violation}",
    eprint = "hep-ph/0604263",
    archivePrefix = "arXiv",
    reportNumber = "ITFA-2006-17, DAMTP-2006-29",
    doi = "10.1088/1126-6708/2006/08/012",
    journal = "JHEP",
    volume = "08",
    pages = "012",
    year = "2006"
}

@article{Tranberg:2006dg,
    author = "Tranberg, Anders and Smit, Jan and Hindmarsh, Mark",
    title = "{Simulations of cold electroweak baryogenesis: Finite time quenches}",
    eprint = "hep-ph/0610096",
    archivePrefix = "arXiv",
    reportNumber = "ITFA-2006-36, DAMTP-2006-88",
    doi = "10.1088/1126-6708/2007/01/034",
    journal = "JHEP",
    volume = "01",
    pages = "034",
    year = "2007"
}

@article{Tranberg:2009de,
    author = "Tranberg, Anders and Hernandez, Andres and Konstandin, Thomas and Schmidt, Michael G.",
    title = "{Cold electroweak baryogenesis with Standard Model CP violation}",
    eprint = "0909.4199",
    archivePrefix = "arXiv",
    primaryClass = "hep-ph",
    reportNumber = "UAB-FT-671, HD-THEP-09-20",
    doi = "10.1016/j.physletb.2010.05.030",
    journal = "Phys. Lett. B",
    volume = "690",
    pages = "207--212",
    year = "2010"
}

@article{Enqvist:2010fd,
    author = "Enqvist, Kari and Stephens, Philip and Taanila, Olli and Tranberg, Anders",
    title = "{Fast Electroweak Symmetry Breaking and Cold Electroweak Baryogenesis}",
    eprint = "1005.0752",
    archivePrefix = "arXiv",
    primaryClass = "astro-ph.CO",
    doi = "10.1088/1475-7516/2010/09/019",
    journal = "JCAP",
    volume = "09",
    pages = "019",
    year = "2010"
}

@article{Tranberg:2010af,
    author = "Tranberg, Anders",
    title = "{Standard Model CP-violation and Cold Electroweak Baryogenesis}",
    eprint = "1009.2358",
    archivePrefix = "arXiv",
    primaryClass = "hep-ph",
    reportNumber = "HIP-2010-22-TH",
    doi = "10.1103/PhysRevD.84.083516",
    journal = "Phys. Rev. D",
    volume = "84",
    pages = "083516",
    year = "2011"
}

@article{Tranberg:2012jp,
    author = "Tranberg, Anders and Wu, Bin",
    title = "{Cold Electroweak Baryogenesis in the Two Higgs-Doublet Model}",
    eprint = "1203.5012",
    archivePrefix = "arXiv",
    primaryClass = "hep-ph",
    reportNumber = "BI-TP-2012-11",
    doi = "10.1007/JHEP07(2012)087",
    journal = "JHEP",
    volume = "07",
    pages = "087",
    year = "2012"
}

@article{Mou:2015aia,
    author = "Mou, Zong-Gang and Saffin, Paul M. and Tranberg, Anders",
    title = "{Cold Baryogenesis from first principles in the Two-Higgs Doublet model with Fermions}",
    eprint = "1505.02692",
    archivePrefix = "arXiv",
    primaryClass = "hep-ph",
    doi = "10.1007/JHEP06(2015)163",
    journal = "JHEP",
    volume = "06",
    pages = "163",
    year = "2015"
}

@article{Mou:2017atl,
    author = "Mou, Zong-Gang and Saffin, Paul M. and Tranberg, Anders",
    title = "{Simulations of Cold Electroweak Baryogenesis: Finding the optimal quench time}",
    eprint = "1703.01781",
    archivePrefix = "arXiv",
    primaryClass = "hep-ph",
    doi = "10.1007/JHEP07(2017)010",
    journal = "JHEP",
    volume = "07",
    pages = "010",
    year = "2017"
}

@article{Mou:2017zwe,
    author = "Mou, Zong-Gang and Saffin, Paul M. and Tranberg, Anders",
    title = "{Simulations of Cold Electroweak Baryogenesis: Hypercharge U(1) and the creation of helical magnetic fields}",
    eprint = "1704.08888",
    archivePrefix = "arXiv",
    primaryClass = "hep-ph",
    doi = "10.1007/JHEP06(2017)075",
    journal = "JHEP",
    volume = "06",
    pages = "075",
    year = "2017"
}

@article{Mou:2017xbo,
    author = "Mou, Zong-Gang and Saffin, Paul M. and Tranberg, Anders",
    title = "{Simulations of Cold Electroweak Baryogenesis: Quench from portal coupling to new singlet field}",
    eprint = "1711.04524",
    archivePrefix = "arXiv",
    primaryClass = "hep-ph",
    doi = "10.1007/JHEP01(2018)103",
    journal = "JHEP",
    volume = "01",
    pages = "103",
    year = "2018"
}

@article{Mou:2018xto,
    author = "Mou, Zong-Gang and Saffin, Paul M. and Tranberg, Anders",
    title = "{Simulations of Cold Electroweak Baryogenesis: Dependence on the source of CP-violation}",
    eprint = "1803.07346",
    archivePrefix = "arXiv",
    primaryClass = "hep-ph",
    doi = "10.1007/JHEP05(2018)197",
    journal = "JHEP",
    volume = "05",
    pages = "197",
    year = "2018"
}

@inproceedings{Moore:2000ara,
    author = "Moore, Guy D.",
    title = "{Do we understand the sphaleron rate?}",
    booktitle = "{4th International Conference on Strong and Electroweak Matter}",
    eprint = "hep-ph/0009161",
    archivePrefix = "arXiv",
    doi = "10.1142/9789812799913_0007",
    pages = "82--94",
    month = "6",
    year = "2000"
}

@article{Dorsch:2021ubz,
    author = "Dorsch, Glauber C. and Huber, Stephan J. and Konstandin, Thomas",
    title = "{On the wall velocity dependence of electroweak baryogenesis}",
    eprint = "2106.06547",
    archivePrefix = "arXiv",
    primaryClass = "hep-ph",
    reportNumber = "DESY-21-089, DESY 21-089",
    doi = "10.1088/1475-7516/2021/08/020",
    journal = "JCAP",
    volume = "08",
    pages = "020",
    year = "2021"
}

@article{Konstandin:2011dr,
    author = "Konstandin, Thomas and Servant, Geraldine",
    title = "{Cosmological Consequences of Nearly Conformal Dynamics at the TeV scale}",
    eprint = "1104.4791",
    archivePrefix = "arXiv",
    primaryClass = "hep-ph",
    doi = "10.1088/1475-7516/2011/12/009",
    journal = "JCAP",
    volume = "12",
    pages = "009",
    year = "2011"
}

@article{vonHarling:2016vhf,
    author = "von Harling, Benedict and Servant, Geraldine",
    title = "{Cosmological evolution of Yukawa couplings: the 5D perspective}",
    eprint = "1612.02447",
    archivePrefix = "arXiv",
    primaryClass = "hep-ph",
    reportNumber = "DESY-16-221",
    doi = "10.1007/JHEP05(2017)077",
    journal = "JHEP",
    volume = "05",
    pages = "077",
    year = "2017"
}

@article{Bruggisser:2018mrt,
    author = "Bruggisser, Sebastian and Von Harling, Benedict and Matsedonskyi, Oleksii and Servant, G{\'e}raldine",
    title = "{Electroweak Phase Transition and Baryogenesis in Composite Higgs Models}",
    eprint = "1804.07314",
    archivePrefix = "arXiv",
    primaryClass = "hep-ph",
    reportNumber = "DESY-17-229",
    doi = "10.1007/JHEP12(2018)099",
    journal = "JHEP",
    volume = "12",
    pages = "099",
    year = "2018"
}

@article{Bruggisser:2022rdm,
    author = "Bruggisser, Sebastian and von Harling, Benedict and Matsedonskyi, Oleksii and Servant, Geraldine",
    title = "{Status of electroweak baryogenesis in minimal composite Higgs}",
    eprint = "2212.11953",
    archivePrefix = "arXiv",
    primaryClass = "hep-ph",
    reportNumber = "DESY-22-209",
    doi = "10.1007/JHEP08(2023)012",
    journal = "JHEP",
    volume = "08",
    pages = "012",
    year = "2023"
}

@article{Bruggisser:2022ofg,
    author = "Bruggisser, Sebastian and von Harling, Benedict and Matsedonskyi, Oleksii and Servant, Geraldine",
    title = "{Dilaton at the LHC: complementary probe of composite Higgs}",
    eprint = "2212.00056",
    archivePrefix = "arXiv",
    primaryClass = "hep-ph",
    reportNumber = "DESY-22-190",
    doi = "10.1007/JHEP05(2023)080",
    journal = "JHEP",
    volume = "05",
    pages = "080",
    year = "2023"
}

@article{Servant:2014bla,
    author = "Servant, Geraldine",
    title = "{Baryogenesis from Strong $CP$ Violation and the QCD Axion}",
    eprint = "1407.0030",
    archivePrefix = "arXiv",
    primaryClass = "hep-ph",
    doi = "10.1103/PhysRevLett.113.171803",
    journal = "Phys. Rev. Lett.",
    volume = "113",
    number = "17",
    pages = "171803",
    year = "2014"
}

@article{Cutting:2018tjt,
    author = "Cutting, Daniel and Hindmarsh, Mark and Weir, David J.",
    title = "{Gravitational waves from vacuum first-order phase transitions: from the envelope to the lattice}",
    eprint = "1802.05712",
    archivePrefix = "arXiv",
    primaryClass = "astro-ph.CO",
    reportNumber = "HIP-2018-4-TH",
    doi = "10.1103/PhysRevD.97.123513",
    journal = "Phys. Rev. D",
    volume = "97",
    number = "12",
    pages = "123513",
    year = "2018"
}

@article{Jinno:2019bxw,
    author = "Jinno, Ryusuke and Konstandin, Thomas and Takimoto, Masahiro",
    title = "{Relativistic bubble collisions\textemdash{}a closer look}",
    eprint = "1906.02588",
    archivePrefix = "arXiv",
    primaryClass = "hep-ph",
    reportNumber = "DESY-19-102, DESY 19-102",
    doi = "10.1088/1475-7516/2019/09/035",
    journal = "JCAP",
    volume = "09",
    pages = "035",
    year = "2019"
}

@article{deVries:2017ncy,
    author = "de Vries, Jordy and Postma, Marieke and van de Vis, Jorinde and White, Graham",
    title = "{Electroweak Baryogenesis and the Standard Model Effective Field Theory}",
    eprint = "1710.04061",
    archivePrefix = "arXiv",
    primaryClass = "hep-ph",
    reportNumber = "Nikhef-2017-044",
    doi = "10.1007/JHEP01(2018)089",
    journal = "JHEP",
    volume = "01",
    pages = "089",
    year = "2018"
}

@article{Grojean:2004xa,
    author = "Grojean, Christophe and Servant, Geraldine and Wells, James D.",
    title = "{First-order electroweak phase transition in the standard model with a low cutoff}",
    eprint = "hep-ph/0407019",
    archivePrefix = "arXiv",
    reportNumber = "SACLAY-T04-084, MCTP-04-37, ANL-HEP-PR-04-63, EFI-04-23",
    doi = "10.1103/PhysRevD.71.036001",
    journal = "Phys. Rev. D",
    volume = "71",
    pages = "036001",
    year = "2005"
}

@article{Delaunay:2007wb,
    author = "Delaunay, Cedric and Grojean, Christophe and Wells, James D.",
    title = "{Dynamics of Non-renormalizable Electroweak Symmetry Breaking}",
    eprint = "0711.2511",
    archivePrefix = "arXiv",
    primaryClass = "hep-ph",
    reportNumber = "CERN-PH-TH-2007-219, MCTP-07-31, SACLAY-T07-141",
    doi = "10.1088/1126-6708/2008/04/029",
    journal = "JHEP",
    volume = "04",
    pages = "029",
    year = "2008"
}

@article{Figueroa:2020rrl,
    author = "Figueroa, Daniel G. and Florio, Adrien and Torrenti, Francisco and Valkenburg, Wessel",
    title = "{The art of simulating the early Universe -- Part I}",
    eprint = "2006.15122",
    archivePrefix = "arXiv",
    primaryClass = "astro-ph.CO",
    doi = "10.1088/1475-7516/2021/04/035",
    journal = "JCAP",
    volume = "04",
    pages = "035",
    year = "2021"
}

@article{Cirigliano:2019vfc,
    author = "Cirigliano, Vincenzo and Crivellin, Andreas and Dekens, Wouter and de Vries, Jordy and Hoferichter, Martin and Mereghetti, Emanuele",
    title = "{CP Violation in Higgs-Gauge Interactions: From Tabletop Experiments to the LHC}",
    eprint = "1903.03625",
    archivePrefix = "arXiv",
    primaryClass = "hep-ph",
    reportNumber = "INT-PUB-19-008, LA-UR-19-22027, PSI-PR-19-01, ZU-TH 08/19, RBRC-1316",
    doi = "10.1103/PhysRevLett.123.051801",
    journal = "Phys. Rev. Lett.",
    volume = "123",
    number = "5",
    pages = "051801",
    year = "2019"
}

@article{Blasi:2022woz,
    author = "Blasi, Simone and Mariotti, Alberto",
    title = "{Domain Walls Seeding the Electroweak Phase Transition}",
    eprint = "2203.16450",
    archivePrefix = "arXiv",
    primaryClass = "hep-ph",
    doi = "10.1103/PhysRevLett.129.261303",
    journal = "Phys. Rev. Lett.",
    volume = "129",
    number = "26",
    pages = "261303",
    year = "2022"
}

@article{Blasi:2023rqi,
    author = "Blasi, Simone and Jinno, Ryusuke and Konstandin, Thomas and Rubira, Henrique and Stomberg, Isak",
    title = "{Gravitational waves from defect-driven phase transitions: domain walls}",
    eprint = "2302.06952",
    archivePrefix = "arXiv",
    primaryClass = "astro-ph.CO",
    doi = "10.1088/1475-7516/2023/10/051",
    journal = "JCAP",
    volume = "10",
    pages = "051",
    year = "2023"
}

@article{Agrawal:2023cgp,
    author = "Agrawal, Prateek and Blasi, Simone and Mariotti, Alberto and Nee, Michael",
    title = "{Electroweak phase transition with a double well done doubly well}",
    eprint = "2312.06749",
    archivePrefix = "arXiv",
    primaryClass = "hep-ph",
    reportNumber = "DESY-23-208",
    doi = "10.1007/JHEP06(2024)089",
    journal = "JHEP",
    volume = "06",
    pages = "089",
    year = "2024"
}

@article{Konstandin:2011ds,
    author = "Konstandin, Thomas and Servant, Geraldine",
    title = "{Natural Cold Baryogenesis from Strongly Interacting Electroweak Symmetry Breaking}",
    eprint = "1104.4793",
    archivePrefix = "arXiv",
    primaryClass = "hep-ph",
    doi = "10.1088/1475-7516/2011/07/024",
    journal = "JCAP",
    volume = "07",
    pages = "024",
    year = "2011"
}

@inproceedings{Moore:1998mh,
    author = "Moore, Guy D.",
    title = "{The Sphaleron rate: Where we stand}",
    booktitle = "{3rd International Conference on Strong and Electroweak Matter}",
    eprint = "hep-ph/9902464",
    archivePrefix = "arXiv",
    pages = "23--33",
    month = "12",
    year = "1998"
}

@article{DOnofrio:2014rug,
    author = "D'Onofrio, Michela and Rummukainen, Kari and Tranberg, Anders",
    title = "{Sphaleron Rate in the Minimal Standard Model}",
    eprint = "1404.3565",
    archivePrefix = "arXiv",
    primaryClass = "hep-ph",
    doi = "10.1103/PhysRevLett.113.141602",
    journal = "Phys. Rev. Lett.",
    volume = "113",
    number = "14",
    pages = "141602",
    year = "2014"
}

@article{Dine:1991ck,
    author = "Dine, Michael and Huet, Patrick and Singleton, Jr., Robert L.",
    title = "{Baryogenesis at the electroweak scale}",
    reportNumber = "SCIPP-91-08",
    doi = "10.1016/0550-3213(92)90113-P",
    journal = "Nucl. Phys. B",
    volume = "375",
    pages = "625--648",
    year = "1992"
}

@article{Lewicki:2020azd,
    author = "Lewicki, Marek and Vaskonen, Ville",
    title = "{Gravitational waves from colliding vacuum bubbles in gauge theories}",
    eprint = "2012.07826",
    archivePrefix = "arXiv",
    primaryClass = "astro-ph.CO",
    doi = "10.1140/epjc/s10052-021-09232-3",
    journal = "Eur. Phys. J. C",
    volume = "81",
    number = "5",
    pages = "437",
    year = "2021",
    note = "[Erratum: Eur.Phys.J.C 81, 1077 (2021)]"
}

@article{Gould:2021dpm,
    author = "Gould, Oliver and Sukuvaara, Satumaaria and Weir, David",
    title = "{Vacuum bubble collisions: From microphysics to gravitational waves}",
    eprint = "2107.05657",
    archivePrefix = "arXiv",
    primaryClass = "astro-ph.CO",
    reportNumber = "HIP-2021-4/TH",
    doi = "10.1103/PhysRevD.104.075039",
    journal = "Phys. Rev. D",
    volume = "104",
    number = "7",
    pages = "075039",
    year = "2021"
}

@article{Bodeker:2017cim,
    author = "Bodeker, Dietrich and Moore, Guy D.",
    title = "{Electroweak Bubble Wall Speed Limit}",
    eprint = "1703.08215",
    archivePrefix = "arXiv",
    primaryClass = "hep-ph",
    doi = "10.1088/1475-7516/2017/05/025",
    journal = "JCAP",
    volume = "05",
    pages = "025",
    year = "2017"
}

@article{Gouttenoire:2021kjv,
    author = "Gouttenoire, Yann and Jinno, Ryusuke and Sala, Filippo",
    title = "{Friction pressure on relativistic bubble walls}",
    eprint = "2112.07686",
    archivePrefix = "arXiv",
    primaryClass = "hep-ph",
    reportNumber = "DESY-21-147, IFT-UAM/CSIC-21-146",
    doi = "10.1007/JHEP05(2022)004",
    journal = "JHEP",
    volume = "05",
    pages = "004",
    year = "2022"
}

@article{Azatov:2023xem,
    author = "Azatov, Aleksandr and Barni, Giulio and Petrossian-Byrne, Rudin and Vanvlasselaer, Miguel",
    title = "{Quantisation across bubble walls and friction}",
    eprint = "2310.06972",
    archivePrefix = "arXiv",
    primaryClass = "hep-ph",
    reportNumber = "SISSA 13/2023/FISI",
    doi = "10.1007/JHEP05(2024)294",
    journal = "JHEP",
    volume = "05",
    pages = "294",
    year = "2024"
}

@article{Azatov:2020ufh,
    author = "Azatov, Aleksandr and Vanvlasselaer, Miguel",
    title = "{Bubble wall velocity: heavy physics effects}",
    eprint = "2010.02590",
    archivePrefix = "arXiv",
    primaryClass = "hep-ph",
    reportNumber = "SISSA 247/2020/FISI",
    doi = "10.1088/1475-7516/2021/01/058",
    journal = "JCAP",
    volume = "01",
    pages = "058",
    year = "2021"
}

@article{Baldes:2024wuz,
    author = "Baldes, Iason and Dichtl, Maximilian and Gouttenoire, Yann and Sala, Filippo",
    title = "{Particle shells from relativistic bubble walls}",
    eprint = "2403.05615",
    archivePrefix = "arXiv",
    primaryClass = "hep-ph",
    doi = "10.1007/JHEP07(2024)231",
    journal = "JHEP",
    volume = "07",
    pages = "231",
    year = "2024"
}

@article{Azatov:2021irb,
    author = "Azatov, Aleksandr and Vanvlasselaer, Miguel and Yin, Wen",
    title = "{Baryogenesis via relativistic bubble walls}",
    eprint = "2106.14913",
    archivePrefix = "arXiv",
    primaryClass = "hep-ph",
    reportNumber = "SISSA 13/2021/FISI TU-1127",
    doi = "10.1007/JHEP10(2021)043",
    journal = "JHEP",
    volume = "10",
    pages = "043",
    year = "2021"
}

@article{Baldes:2021vyz,
    author = "Baldes, Iason and Blasi, Simone and Mariotti, Alberto and Sevrin, Alexander and Turbang, Kevin",
    title = "{Baryogenesis via relativistic bubble expansion}",
    eprint = "2106.15602",
    archivePrefix = "arXiv",
    primaryClass = "hep-ph",
    reportNumber = "ULB-TH/21-09",
    doi = "10.1103/PhysRevD.104.115029",
    journal = "Phys. Rev. D",
    volume = "104",
    number = "11",
    pages = "115029",
    year = "2021"
}

@article{Cataldi:2024pgt,
    author = "Cataldi, Martina and Shakya, Bibhushan",
    title = "{Leptogenesis via bubble collisions}",
    eprint = "2407.16747",
    archivePrefix = "arXiv",
    primaryClass = "hep-ph",
    reportNumber = "DESY-24-110",
    doi = "10.1088/1475-7516/2024/11/047",
    journal = "JCAP",
    volume = "11",
    pages = "047",
    year = "2024"
}

@article{Bodeker:2009qy,
    author = "Bodeker, Dietrich and Moore, Guy D.",
    title = "{Can electroweak bubble walls run away?}",
    eprint = "0903.4099",
    archivePrefix = "arXiv",
    primaryClass = "hep-ph",
    doi = "10.1088/1475-7516/2009/05/009",
    journal = "JCAP",
    volume = "05",
    pages = "009",
    year = "2009"
}

@article{Vachaspati:2024vbw,
    author = "Vachaspati, Tanmay and Brandenburg, Axel",
    title = "{Spectra of magnetic fields from electroweak symmetry breaking}",
    eprint = "2412.00641",
    archivePrefix = "arXiv",
    primaryClass = "astro-ph.CO",
    reportNumber = "NORDITA-2024-047",
    doi = "10.1103/PhysRevD.111.043541",
    journal = "Phys. Rev. D",
    volume = "111",
    number = "4",
    pages = "043541",
    year = "2025"
}

@article{Di:2020kbw,
    author = "Di, Yuefeng and Wang, Jialong and Zhou, Ruiyu and Bian, Ligong and Cai, Rong-Gen and Liu, Jing",
    title = "{Magnetic Field and Gravitational Waves from the First-Order Phase Transition}",
    eprint = "2012.15625",
    archivePrefix = "arXiv",
    primaryClass = "astro-ph.CO",
    doi = "10.1103/PhysRevLett.126.251102",
    journal = "Phys. Rev. Lett.",
    volume = "126",
    number = "25",
    pages = "251102",
    year = "2021"
}

@article{Di:2024gsl,
    author = "Di, Yuefeng and Bian, Ligong and Cai, Rong-Gen",
    title = "{Baryogenesis Induced by Magnetic Field Effects During the Electroweak Phase Transition}",
    eprint = "2409.16124",
    archivePrefix = "arXiv",
    primaryClass = "hep-ph",
    month = "9",
    year = "2024"
}

@article{Di:2025ncl,
    author = "Di, Yuefeng and Bian, Ligong and Cai, Rong-Gen",
    title = "{Impact of Primordial Magnetic Fields on the First-Order Electroweak Phase Transition}",
    eprint = "2508.07416",
    archivePrefix = "arXiv",
    primaryClass = "hep-ph",
    month = "8",
    year = "2025"
}

@article{Dichtl:2023xqd,
    author = "Dichtl, Maximilian and Nava, Jacopo and Pascoli, Silvia and Sala, Filippo",
    title = "{Baryogenesis and leptogenesis from supercooled confinement}",
    eprint = "2312.09282",
    archivePrefix = "arXiv",
    primaryClass = "hep-ph",
    doi = "10.1007/JHEP02(2024)059",
    journal = "JHEP",
    volume = "02",
    pages = "059",
    year = "2024"
}

\end{document}